\documentclass[twocolumn,pre,showpacs,floatfix,eqsecnum]{revtex4}
\usepackage{bm}
\usepackage{array}

\newcommand{\gtapprox}{{\raise.3ex\hbox{$>$\kern-.75em\lower1ex\hbox{$\sim$}}}}

\usepackage[dvips]{epsfig}

\usepackage{amsmath}
\usepackage{amsfonts}

\newcommand{\kbar}{{\overline k}}
\newcommand{\hatq}{{\hat q}}
\newcommand{\Ha}{{\cal H}}
\newcommand{\Ft}{{\tilde F}}
\newcommand{\Ftv}{\tilde{\mathbf{F}}}
\newcommand{\sigmat}{{\tilde \sigma}}
\newcommand{\sigmab}{{\overline \sigma}}
\newcommand{\ellv}{{\mathbf \ell}}
\newcommand{\Rv}{{\mathbf R}}
\newcommand{\rv}{{\mathbf r}}
\newcommand{\ev}{{\mathbf e}}
\newcommand{\uv}{{\mathbf u}}
\newcommand{\qv}{{\mathbf q}}
\newcommand{\xv}{{\mathbf x}}
\newcommand{\fv}{{\mathbf f}}
\newcommand{\Fv}{{\mathbf F}}
\newcommand{\bfx}{{\bf x}}
\newcommand{\bfu}{{\bf u}}

\newcommand{\begineq}[1]{\begin{equation}\label{#1}}
\newcommand{\eqend}{\end{equation}}

\newcommand{\xsep}{{\left| \xv \right|}}
\newcommand{\gfunc}{{\mathcal{G}(\mathbf{x})}}

\newcommand{\calF}{{\mathcal{F}}}
\newcommand{\calK}{{\mathcal{K}}}

\begin{document}
\title{Nonaffine Correlations in Random Elastic Media}
\author{B.A. DiDonna}
\affiliation{Institute for Mathematics and its Applications,
University of Minnesota, Minneapolis, MN 55455-0436, USA}
\author{T.C. Lubensky}
\affiliation{Department of Physics and Astronomy, University of
Pennsylvania, Philadelphia, PA 19104-6396, USA}

\begin{abstract}
Materials characterized by spatially homogeneous elastic moduli
undergo affine distortions when subjected to external stress at
their boundaries, i.e., their displacements $\uv ( \xv )$ from a
uniform reference state grow linearly with position $\xv$, and
their strains are spatially constant.  Many materials, including
all macroscopically isotropic amorphous ones, have elastic moduli
that vary randomly with position, and they necessarily undergo
nonaffine distortions in response to external stress.  We study
general aspects of nonaffine response and correlation using
analytic calculations and numerical simulations. We define
nonaffine displacements $\uv' ( \xv)$ as the difference between
$\uv ( \xv)$ and affine displacements, and we investigate the
nonaffinity correlation function $\gfunc = \langle [\uv'( \xv ) -
\uv' ( 0 )]^2 \rangle$ and related functions. We introduce four
model random systems with random elastic moduli induced by locally
random spring constants (none of which are infinite), by random
coordination number, by random stress, or by any combination of
these. We show analytically and numerically that $\gfunc$ scales
as $A |\xv|^{-(d-2)}$ where the amplitude $A$ is proportional to
the variance of local elastic moduli regardless of the origin of
their randomness. We show that the driving force for nonaffine
displacements is a spatial derivative of the random elastic
constant tensor times the constant affine strain.  Random stress
by itself does not drive nonaffine response, though the randomness
in elastic moduli it may generate does. We study models with both
short and long-range correlations in random elastic moduli.
\end{abstract}


\maketitle

\section{\label{intro}Introduction}

In the classical theory of elasticity
\cite{Landau-elasticity,BornHua1954,Love1944,ChaikinLub95}, an
elastic material is viewed as a spatially homogeneous medium
characterized by a spatially constant elastic-modulus tensor
$K_{ijkl}$. When such a medium is subjected to uniform stresses at
its boundaries, it will undergo a homogeneous deformation with a
constant strain. Such homogeneous deformations are called affine.
This picture of affine strain is generally valid at length scales
large compared to any characteristic inhomogeneities:
displacements averaged over a sufficiently large volume are affine
(at least in dimensions greater than two). It applies not only to
regular periodic crystals, but also to polycrystalline materials
like a typical bar of steel. At more microscopic scales, however,
individual particles in an elastic medium do not necessarily
follow trajectories defined by uniform strain in response to
external stress: they undergo nonaffine rather than affine
displacements. The only systems that are guaranteed to exhibit
affine distortions at the microscopic scale are periodic solids
with a single atom per unit cell.  Atoms within a multi-atom unit
cell of a periodic solid will in general undergo nonaffine
distortions \cite{JaricMoh1988}, and atoms in random and amorphous
solids will certainly undergo nonaffine distortions. Such
distortions can lead to substantial corrections to the Born-Huang
\cite{BornHua1954} expression for macroscopic elastic moduli.

Research on fragile
\cite{Durian1995,Durian1997,LangerLiu1997,TewariLiu1999,EvansCates2000},
granular \cite{JaegerBeh1996,HalseyMehta2003}, crosslinked
polymeric
\cite{RubinsteinPan1997,RubinsteinPan2002,GlattingRei1997,Everaers1998,SvaneborgEve2004,SommerLay2002},
and biological materials
\cite{MacKintoshJan1995,HeadMac2003c,HeadMac2003a,HeadMac2003b,WilhelmFre2003},
particularly in small samples, has sparked a renewed interest in
the nature of nonaffine response and its ramifications.  Liu and
Langer \cite{LangerLiu1997} introduced various measures of
nonaffinity, in particular the mean-square deviation from affinity
of individual particles in model foams subjected to shear. Tanguy
{\it et al.} \cite{TanguyBar2002} in their simulation of amorphous
systems of Lennard-Jones beads found substantial nonaffine
response and a resultant size-dependence to the macroscopic
elastic moduli. Lemaitre and Maloney \cite{LemaitreMal2005} relate
nonaffinity to a random force field induced by an initial affine
response. Head {\it et al.}
\cite{HeadMac2003c,HeadMac2003a,HeadMac2003b} studied models of
crosslinked semi-flexible rods in two-dimensions and found two
types of behavior depending on the density of rods. In dense
systems, the response is close to affine and is dominated by rod
compression, whereas in more dilute systems, the response is
strongly nonaffine and dominated by rod bending.

The recent work discussed above provides valuable insight into the
nature of nonaffine response. It does not, however, provide a
general framework in which to describe it. In this paper, we
provide a such a framework for describing the long-wavelength
properties of nonaffinity, and we verify its validity with
numerical calculations of these properties on a number of
zero-temperature central-force lattice models specifically
designed to demonstrate our ideas.  Our hope is that this
framework will prove a useful tool for studying more realistic
models of amorphous glasses, granular material, and jammed
systems, particularly at zero temperature just above the jamming
transition \cite{LiuNag1998,OhernNag2001,OhernNag2003}. We are
currently applying them to jammed systems \cite{VernonLub2005} and
to networks of semi-flexible polymers \cite{DidonnaLev2005}.

Though nonaffinity concerns the displacement of individual
particles at the microscopic scale, we show that general aspects
of nonaffine response in random and amorphous systems can be
described in terms of a continuum elastic model characterized by a
local elastic-modulus tensor $K_{ijkl}(\xv)$ at point $\xv$,
consisting of a spatially uniform average part $K_{ijkl}$ and a
locally fluctuating part $\delta K_{ijkl}(\xv)$, and possibly a
local stress tensor $\tilde{\sigma}_{ij}( \xv)$ with vanishing
mean. We show that under stress leading to a macroscopic strain
$\gamma_{ij}$, the random part of the elastic-modulus tensor, in
conjunction with the strain $\gamma_{ij}$, acts as a source of
nonaffine displacement $u'_i(\xv)$ proportional to $\partial_j
\delta K_{ijkl}\gamma_{kl}$. For small $\delta K_{ijkl}$ and
$\gamma_{ij}$, the Fourier transform of the correlation function
$G_{ij} ( \xv,0)$ of the displacement $\uv'( \xv)$ can be
expressed schematically as $\gamma^2 \Delta^K(\qv)/(q^2 K^2)$
where $\Delta^K(\qv)$ represents the Fourier transform of relevant
components of the correlation function of the random part of the
elastic-modulus tensor and $K$ represents the average
elastic-modulus tensor. At length scales large compared to the
correlation length $\xi$ of the random elastic modulus,
$\Delta^K(\qv)$ is a constant $\Delta^K$, and the nonaffinity
correlation function in $d$ dimensions scales as $(\Delta^K /K^2)
\gamma^2 |\xv|^{-(d-2)}$, which exhibits, in particular, a
logarithmic divergence in two dimensions; at length scales smaller
than $\xi$, $\Delta^K ( \qv) \sim q^{-\phi}$, where $\phi$ can be
viewed as a critical exponent, and the nonaffinity correlation
function scales as $|\xv|^{\phi+2 - d}$ for $\phi< d$. For
simplicity, we focus on zero-temperature systems.  Our analytic
approach is, however, easily generalized to nonzero temperature in
systems with unbreakable bonds.  At nonzero temperature, the
dominant, long-distance behavior of nonaffinity correlation
functions is the same as at zero temperature.

Our numerical studies were carried out on systems composed of
sites either on regular periodic lattices or on random lattices
constructed by sampling a Lennard-Jones liquid and connecting
nearest-neighbor sites with unbreakable central force springs.  We
allowed the spring constants of the springs, their preferred
lengths, or both to vary randomly. The local elastic modulus at a
particular site in these models depends on the strength and length
of springs connected to that site as well as on the number of
springs connected to it. Thus, a periodic lattice with random
spring constants and an amorphous lattice with random site
coordination numbers both have a random local elastic constant.
Their nonaffinity correlation function should, therefore, exhibit
similar behavior, as our calculations and simulations verify. It
is important to note that macroscopically isotropic systems are
always amorphous and, therefore, always have a random
elastic-modulus tensor and exhibit nonaffine response.
For simplicity, we do not consider systems in which any spring is
infinitely rigid (i.e., has an infinite spring constant).  With
appropriate coarse graining of $\delta K_{ijkl} (\xv)$, however,
our primary analytical results are expected to apply to this more
general case.

The outline of this paper is as follows.  In Sec.\
\ref{Models_def}, we derive familiar formulae for the elastic
energy of central-force lattices and introduce our continuum
model, giving special attention to the nature of random stresses.
In Sec.\ \ref{strains_nonaff}, we use the continuum model to
calculate nonaffine response functions in different dimensions for
systems with random elastic moduli with both short- and long-range
correlations and with random stress tensors relative to a uniform
state, and we calculate the correlation function of local
rotations induced by nonaffine distortions. In Sec.\
\ref{Num_min}, we present numerical results for the four model
systems we consider: periodic lattices with random elastic
constants without (Model A) and with (Model B) random stress, and
amorphous lattices with random elastic constants without (Model C)
and with (Model D) random stress. Four appendices present
calculational details:  Appendix \ref{App:A} derives the
independent components of the $8$th rank modulus correlator in an
isotropic medium, App.\ \ref{App:B} calculates the general form of
the nonaffinity correlation function as a function of wavevector,
App.\ \ref{App:C} calculates the asymptotic forms as a function of
separation $\xv$ of the nonaffinity correlation function, and
App.\ \ref{App:D} calculates the correlation function of local
vorticity.

\section{\label{Models_def}Models and Definitions}

\subsection{Notation and Model Energy}
\label{model_en}

We consider model elastic networks in which particles occupy sites
on periodic or random lattices in their force-free equilibrium
state. Thus, particle $\ellv$ is at lattice position $\Rv_{\ellv
0}$ in equilibrium. When the lattices are distorted, particle
$\ellv$ undergoes a displacement $\uv_{\ellv}$ to a new position
\begin{equation}
\Rv_{\ellv} = \Rv_{\ellv 0} + \uv_{\ellv} .
\end{equation}
We will refer to the equilibrium lattice, with lattice positions
$\Rv_{\ellv 0}$, as the {\bf reference lattice} or {\bf reference
space}, and the space into which the lattice is distorted via the
displacements $\uv_{\ellv}$ as the {\bf target space}. Pairs of
particles $\ellv$ and $\ellv'$ are connected by unbreakable
central-force springs on the bond $b \equiv <\ellv',\ellv>$. The
coordination number of each particle (or site) is equal to the
number of particles (or sites) to which it is connected by bonds.
The potential energy, $V_b(R_b)$, of the spring on bond $b$
depends only on the magnitude,
\begin{equation}
R_b = |\Rv_{\ellv'}-\Rv_{\ellv}| ,
\end{equation}
of the vector connecting particles $\ellv$ and $\ellv'$. The total
potential energy is thus
\begin{equation}
U_T =  \sum_b V_b ( R_b) \equiv \frac{1}{2}\sum_{\ellv, \ellv'}
V_{<\ellv',\ellv>}(|\Rv_{\ellv'} - \Rv_{\ellv}|) .
\end{equation}
We will consider anharmonic potentials
\begin{equation}
V_b = \frac{1}{2} k_b (R_b - R_{b R})^2 + \frac{1}{4} g_b (R_b -
R_{b R})^4 ,
\end{equation}
with both harmonic and quartic components, where $R_{b R}$ is the
rest length of bond $b$. We assume that both $k_b$ and $g_b$ are
finite. The harmonic limit is obtained when the quartic
coefficient $g_b$ vanishes, in which case, $k_b$ is the harmonic
spring constant.

We will only study systems in which there is an equilibrium
reference state with particle positions $\{\Rv_{\ellv 0} \}$ in
which the force on each site is zero.  The length $R_{b0} \equiv
|\Rv_{\ellv' 0} - \Rv_{\ellv 0}|$ of each bond $b$ in this
configuration does not have to coincide with its rest length
$R_{bR}$.  As we shall see in more detail shortly, it is possible
to have the total force on every site be zero but still have
nonzero forces on each bond.

The potential energy of the lattice can be expanded in terms of
the discrete lattice nonlinear strain \cite{BornHua1954},
\begin{equation}
v_b = \frac{1}{2} (R_b^2 - R_{b 0}^2) = \Rv_{b 0}\cdot \Delta
\uv_b + \frac{1}{2} (\Delta \uv_b \cdot \Delta \uv_b )
\end{equation}
relative to the reference state, where $\Delta \uv_b =
\uv_{\ellv'} - \uv_{\ell}$. The discrete strain variable, $v_b$,
is by construction invariant with respect to rigid rotations of
the sample, i.e., it is invariant under $R_{\ellv i} \rightarrow
U_{ij} R_{\ellv j}$, where $U_{ij}$ is any $\ellv$-independent
rotation matrix. To second order in $v_b$ in an expansion about a
reference lattice with lattice sites $\Rv_{\ellv 0}$, the
potential energy is \cite{BornHua1954}
\begin{equation}
\Delta U_T  =  \sum_b R_{b0}^{-1} \Ft(b) v_b + \frac{1}{2}\sum_b
R_{b0}^{-2} k(b) v_b^2 ,
\label{eq:U_T}
\end{equation}
where $\Ft(b) = |\Ftv (b) |$ is the magnitude of the force,
\begin{equation}
\Ftv (b) = - V_b^{\prime}(R_{b0})\Rv_{b0}/R_{b0} ,
\end{equation}
acting on bond $b$ and
\begin{equation}
k(b) =  V_{b0}^{\prime\prime}(R_{b0}) -
R_{b0}^{-1}V_{b0}^{\prime}(R_{b0})
\end{equation}
is the effective spring constant of bond $b$, which reduces to
$k_b$ when $R_{b0} = R_{bR}$.
$k(b)$ is never infinite because we
we assume $k_b$ and $g_b$ are finite.
The equilibrium bond-length $R_{b0}$ for each bond is determined
by the condition that the total force at each site $\ellv$ vanish
at $\uv_{\ellv} = 0$:
\begin{equation}
F_i (\ellv) = - \left. \frac{\partial \Delta U_T}{\partial
u_{\ellv i}}\right|_{\uv_{\ellv} = 0}= \sum_{\ellv'}
\Ft_i(<\ellv',\ellv>) .
\label{eq:force_eq}
\end{equation}
This equilibrium condition only requires that the total force on
each site, arising from all of the springs attached to it, be zero
\cite{Alexander1998}.  It does not require that the force $\Ftv
(b)$ be equal to zero on every bond $b$.

In equilibrium, when Eq.\ (\ref{eq:force_eq}) is satisfied, the
part of $v_b$ linear in $\Delta \uv_b$ disappears from $\Delta
U_T$.  In this case, it is customary to express $\Delta U_T$ to
harmonic order in $\Delta \uv_b$:
\begin{equation}
\Delta U_T^{\rm har} = \frac{1}{2} \sum_b [V_b^{\prime\prime}
e_{b0i} e_{b0j} + R_{b0}^{-1} V_b' ( \delta_{ij} - e_{b0i}
e_{b0j})] \Delta u_{bi} \Delta u_{bj} , \label{eq:harmU_T}
\end{equation}
where $e_{b0i} = R_{b0i}/R_{b0}$ is the unit vector directed along
bond $b$.  Thus the harmonic potential on each bond decomposes
into a parallel part, proportional to $V_b^{\prime\prime}$,
directed along the bond and a transverse part, proportional to
$R_{b0}^{-1} V_b'$, directed perpendicular to the bond.  The
transverse part vanishes when the force on the bond vanishes.

The harmonic energy $\Delta U_T^{\rm har}$ does not preserve the
invariance with respect to arbitrary rotations of the full
nonlinear strain energy $\Delta U_T$ of Eq.\ (\ref{eq:U_T}), under
which
\begin{equation}
\Delta u_{bi} \rightarrow \Delta u'_{bi} - (U_{ij} - \delta_{ij})
R_{b0j} + U_{ij} \Delta u_{bj} ,
\end{equation}
where $U_{ij}$ is a rotation matrix.  It does, however preserve
this invariance up to order $\theta^2$ but not order $\theta^2
\Delta u_b$ and $\theta (\Delta u_b)^2$, where $\theta$ is a
rotation angle. For small $\bm{\theta}$,
\begin{equation}
\Delta \uv'_b = \Delta \uv_b + \bm{\theta} \times \Rv_{b0} +
O(\theta^2, \theta \Delta u_b ) ,
\label{eq:uv'}
\end{equation}
and $\ev_{b0} \cdot \Delta \uv'_b = \ev_{b0} \cdot \Delta \uv_b +
O(\theta^2, \theta \Delta u_b)$.   Thus, the part of the harmonic
energy arising from the $k(b)$ term in Eq.\ (\ref{eq:U_T}) is
invariant to the order stated above.  The invariance of the force
term of Eq.\ (\ref{eq:U_T}) is more subtle.  Under the above
transformation of Eq.\ (\ref{eq:uv'}), $(\Delta u'_b)^2= (\Delta
u_b)^2 + 2\bm{\theta}\times\Rv_b\cdot\Delta \uv_b +
(\bm{\theta}\times\Rv_b)^2 + O(\theta^2\Delta u_b,\theta (\Delta
u_b)^2)$, and it would seem that there are terms of order
$\theta$, and $\theta^2$ in $\Delta U_T^{\rm har}$.  These terms
vanish, however, upon summation over $\ellv$ and $\ellv'$ because
of the equilibrium force condition of Eq.\ (\ref{eq:force_eq}).
Thus, the full $\Delta U_T^{\rm har}$ is invariant under rotations
up to order $\theta^2$.

\subsection{Definition of Models}
We will consider the following simple models of random lattices.
\\
\newline
\noindent {\bf Model A: Random, zero-force bonds on a periodic
lattice.} In this model, all sites lie on a periodic Bravais
lattice with all bond lengths constant and equal to $R_{b0}$, and
the rest length $R_{bR}$ of each bond is equal to $R_{b0}$.  The
force $\Ftv(b)$ on each bond is zero, but the spring constant
$k_b$ and other properties of the potential $V_b$ can vary from
site to site. Each lattice site has the same coordination number.
\\
\newline
{\bf Model B: Random, finite-force bonds on an originally periodic
lattice.} In this model, sites are originally on a regular
periodic lattice, but rest bond lengths $R_{bR}$ are not equal to
the initial constant bond length on the lattice.  Sites in this
model will relax to positions $R_{\ellv 0}$ with bond lengths
$R_{b 0} = |\Rv_{\ellv' 0} - \Rv_{\ellv 0}|$ such that the force
$\Fv ( \ellv )$ at each site $\ellv$ is zero but the force $\Ftv (
b)$ exerted by each bond $b$ is in general not.  This model has
random stresses and, as we shall see, random elastic moduli as
well. The bond vectors $\Rv_{b 0}$ and spring constant $k_b$ are
random variables, but the coordination number of each site is not.
Random stresses in an originally periodic lattice necessarily
induce randomness in the elastic moduli relative to the relaxed
lattices with zero force at each site.
\\
\newline
{\bf Model C: Random, zero-force bonds on a random lattice.} In
this model, lattice sites are at random positions and have random
coordination numbers.  The equilibrium length $R_{b0}$ varies from
bond to bond.  The rest length $R_{bR}$ of each bond is equal to
its equilibrium length so that the force $\Ftv(b)$ of each bond is
zero. This model, which is meant to describe an amorphous
material, is macroscopically but not microscopically homogeneous
and isotropic.
\\
\newline
{\bf Model D: Random finite-force bonds on a random lattice.} This
is the most general model, and it is the one that provides the
best description of glassy and random granular materials. In it,
the rest lengths $R_{bR}$, the spring constants $k_b$, and the
coordination number are all random variables.  Like Model C, this
model describes macroscopically isotropic and homogeneous
amorphous material.
\\
\newline

Though Models A, B, and C can be viewed as subsets of the most
general model D, we find it useful to treat them as distinct
models because they each isolate separate causes of randomness in
the local elastic modulus or stress. One of our goals, for
example, is to show analytically and numerically that the
non-affinity correlations arising from structural randomness in
models C and D have exactly the same form as those arising from
the more controlled periodic models A and B. Another is to study
the different effects of random elastic moduli and random stress.

In all of these models the random elastic-modulus tensor can in
principle exhibit either short- or long-range correlations in
space. To investigate the effects of such long-range correlations,
we explicitly construct spring constant distributions with
long-range correlations in model A.  We will also find evidence of
long-range correlations in model C when the reference lattice has
correlated crystalline domains.

\subsection{Continuum Models}
In the continuum limit, when spatial variations are slow on a
scale set by the lattice spacing, the equilibrium lattice
positions become continuous positions $\xv$ in the reference
space: $\Rv_{\ellv 0} \rightarrow \xv$; and the target-space
position and displacement vectors become functions of $\xv$:
$\Rv_{\ellv} \rightarrow \Rv(\xv)$ and $\uv_{\ellv} \rightarrow
\uv(\xv)$. In this limit, the lattice strain $v_b$ becomes
\begin{equation}
v_b \approx R_{bi}^0 R_{bj}^0 u_{ij}( \xv) ,
\end{equation}
where
\begin{equation}
u_{ij} (\xv) = \frac{1}{2}(\partial_i u_j + \partial_j u_i +
\partial_i \uv \cdot \partial_j \uv )
\end{equation}
is the full Green-Saint Venant Lagrangian nonlinear strain
\cite{Love1944,Landau-elasticity,ChaikinLub95}, which is invariant
with respect to rigid rotations in the target space [i.e., with
respect to rigid rotations of $\Rv(\xv)$]. Sums over lattice sites
of the form $\sum_{\ellv} S(\ellv)$, for any function $S(\ellv)$,
can be replaced by integrals $\int d^d x S(\xv)/v(\xv)$ where
$v(\xv)$ is the volume of the Voronoi cell centered at position
$\xv = \Rv_{\ellv 0}$. The continuum energy is then
\begin{equation}
\Ha =  \int d^d x \left[ \frac{1}{2} K_{ijkl} ( \xv ) u_{ij} ( \xv
) u_{kl} ( \xv ) + \sigmat_{ij} (\xv ) u_{ij} (\xv) \right],
\label{eq:Ha}
\end{equation}
where
\begin{equation}
\sigmat_{ij} (\xv) = -\frac{1}{2v(\xv) }\sum_{\ellv'} {\tilde F}_i
(b) R_{b0j}|_{b=<\ellv',\ellv>} \label{eq:stress_tensor_1}
\end{equation}
is a local symmetric stress tensor at $\xv$ where the sum over
$\ellv'$ is over all bonds with one end at $\ellv$ and
\begin{equation}
K_{ijkl} ( \xv ) = \frac{1}{2 v(\xv)} \sum_{\ellv'} k(b)
R_{b0}^{-2} R_{b 0 i} R_{b 0 j} R_{b 0 k} R_{b 0 l}|_{b =
<\ellv',\ellv>}
\label{eq:local_el_modulus}
\end{equation}
is the local elastic-modulus tensor \cite{pressure}. Because it
depends only on the full nonlinear strain $u_{ij} ( \xv )$, the
continuum energy $\Ha$ of Eq.\ (\ref{eq:Ha}) is invariant with
respect to rigid rotations in the target space. This is a direct
result of the fact that we consider only internal forces between
particles. The stress tensor $\sigmat_{ij} ( \xv)$ is generated by
these internal forces, and as a result, it multiplies $u_{ij}$ in
$\Ha$. It is necessarily symmetric, and it transforms like a
tensor in the reference space. (It is not, however, the second
Piola-Kirchoff tensor \cite{MarsdenHug1968}, $\sigma_{ij}^{II} =
\delta \Ha/\delta u_{ij} ( \xv )= K_{ijkl} u_{kl} + \sigmat_{ij}$,
which also transforms in this way.) External stresses, on the
other hand, specify a force direction in the target space and
couple to the linear part of the strain.

Since $K_{ijkl}(\xv)$ in Eq.\ (\ref{eq:local_el_modulus}) arises
from central forces on bonds, it and its average over randomness
obey the Cauchy relations \cite{Love1944,BornHua1954},
$K_{ijkl}(\xv) = K_{ikjl}(\xv)=K_{iljk}(\xv)$, in addition to the
more general symmetry relations, $K_{ijkl}(\xv) = K_{jikl}(\xv) =
K_{ijlk}(\xv) = K_{klij}(\xv)$. The Cauchy relations reduce the
number of independent elastic moduli in the average modulus
$K_{ijkl} = \langle K_{ijkl} ( \xv ) \rangle$ below the maximum
number permitted for a given point-group symmetry (for the lowest
symmetry, from 21 to 15). In particular, they reduce the number of
independent moduli in isotropic and hexagonal systems from two to
one, setting the Lam\'{e} coefficients $\lambda$ and $\mu$ equal
to each other. In our analytical calculations, we will, however,
treat $\lambda$ and $\mu$ as independent.  The Cauchy limit is
easily obtained by setting $\lambda = \mu$.

The stress tensor $\sigmat_{ij} (\xv)$ is generated entirely by
internal forces on bonds. The elastic-modulus tensor
$K_{ijkl}(\xv)$ depends on the local effective spring constant
$k(b)$, the length and direction of the bond vectors $\Rv_{b0}$,
and the site coordination number; and it will be a random function
of position if any of these variables are random functions of
position.  Thus $K_{ijkl}( \xv )$ is a random function of position
in Models A to D.  The stress tensor $\sigmat_{ij}(\xv)$ is
nonzero only if the bond forces are nonzero.  It is thus a random
function of position only in Models B and D.

We require that the continuum limit of our lattice models be in
mechanical equilibrium when $\uv(\xv) = 0$.  This means that the
linear variation of $\Ha$ with respect to $\uv(\xv)$ must be zero,
i.e., that
\begin{equation}
\delta \Ha = \int d^d x \sigmat_{ij} ( \xv )
\partial_j \delta u_i ( \xv ) = 0
\end{equation}
for any $\delta u_i ( \xv )$. $\delta u_i ( \xv)$ can be
decomposed into a constant strain part and a part whose average
strain vanishes: $ \delta u_i ( \xv ) = \delta \gamma_{ij} x_j +
\delta \uv'( \xv )$ where $\int d^d \partial_j\delta u'_i ( \xv )
= 0$. Equilibrium with respect to variations in $\gamma_{ij}$
implies that the spatial average of $\sigmat_{ij}$ is zero.
Equilibrium with respect to $\delta \uv' ( \xv )$ implies that
when $\xv$ is in the interior of the sample,
\begin{equation}
f_i ( \xv ) = \partial_j \sigmat_{ij} ( \xv ) = \partial_j
\sigmat_{ji} = 0 ,
\end{equation}
where $\fv$ is the force density that is a vector in the target
space. In addition, $\int d S_j \sigmat_{ij} ( \xv) \delta u'_j (
\xv)= 0$ for any $\delta u_j ( \xv )$, where the integral is over
the surface of the sample, implying that $\sigmat_{ij}(\xv_B) = 0$
for points $\xv_B$ on the surface.

Thus, we see that equilibrium conditions in the reference space
impose stringent constraints on the random stress tensor
$\sigmat_{ij} ( \xv )$: its spatial average must be zero, its
values on sample surfaces must be zero, and it must be purely
transverse, i.e., it must have no longitudinal components parallel
to the gradient operator. Though the linear part of $u_{ij}$ does
not contribute to the stress term in $\Ha$, the nonlinear part
still does, and $\Ha$ can be written as
\begin{eqnarray}
\Ha & = & \frac{1}{2}\int d^d x \left[  K_{ijkl} ( \xv )
u_{ij} ( \xv ) u_{kl} ( \xv ) \right .\nonumber \\
& & \left.+ \sigmat_{ij}(\xv)
\partial_i u_k ( \xv )
\partial_j u_k ( \xv) \right] .
\label{eq:stress3}
\end{eqnarray}
Because of the constraints on $\sigmat_{ij}$, this free energy is
identical to that of Eq.\ (\ref{eq:Ha}).  It is invariant with
respect to rotations in the target space even though it is written
so that the explicit dependence on the rotationally invariant
strain is not so evident \cite{rot_inv}.

As we have seen, the spatial average of $\sigmat_{ij} ( \xv )$ is
zero;  it only has a random fluctuating part in models we
consider.  The elastic-modulus tensor $K_{ijkl} ( \xv )$, on the
other hand, has an average part and a random part with zero mean:
\begin{equation}
K_{ijkl} (\xv) = K_{ijkl} + \delta K_{ijkl} (\xv) .
\end{equation}
We will view both $\sigmat_{ij} ( \xv )$ and $\delta K_{ijkl} (
\xv )$ as quenched random variables with zero mean.

\section{\label{strains_nonaff}Strains and nonaffinity}

\begin{figure}
\centerline{\includegraphics{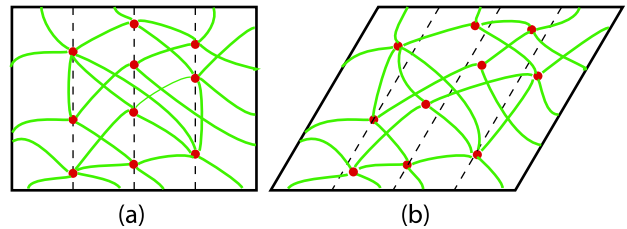}}
\caption{Sheared elastic medium with nonaffine displacements.
(a) Unsheared reference state. (b) Sheared state with nonaffine
displacements. Under affine distortions, points on on the vertical
dotted lines in (a) would map to points on the slanted dotted
lines parallel to the left and right of boundaries of the sheared
sample in (b); under nonaffine distortion, they do not. [Color
online]}
\label{fig:affinity1}
\end{figure}

Consider a reference elastic body in the shape of a regular
parallelepiped.  When such a body is subjected to stresses that
are uniform across each of its faces, it will undergo a strain
deformation in which its boundary sites at positions $\xv_B$
distort to new positions
\begin{equation}
R_i ( \xv _B) = \Lambda_{ij} x_{Bj} , \label{boundary_1}
\end{equation}
where $\Lambda_{ij}$ is the deformation gradient tensor
\cite{MarsdenHug1968}. If the medium is spatially homogeneous,
then $\Lambda_{ij}\equiv \delta_{ij} + \gamma_{ij}$ determines the
displacements of all points in the medium: $R_i (\xv) =
\Lambda_{ij} x_j$ or $u_i ( \xv ) = \gamma_{ij} x_j$.  Such a
distortion is called affine.  In inhomogeneous elastic media,
there will be local deviations from affinity [Fig.\
\ref{fig:affinity1}] described by a displacement variable $\uv' (
\xv )$ defined via
\begin{equation}
R_i ( \xv ) =\Lambda_{ij} x_j  + u'_i ( \xv )
\end{equation}
or, equivalently,
\begin{eqnarray}
u_i ( \xv ) & = & \gamma_{ij}x_j + u'_i ( \xv ) \\
u_{ij}(\xv) & \approx & \gamma_{ij}^S x_j + (\partial_i u'_j +
\partial_j u'_i \nonumber \\
& & +\gamma_{ip}\partial_j u'_p + \gamma_{jp}\partial_i u'_p) /2,
\label{eq:u_displace}
\end{eqnarray}
where the final equation contains only terms up to linear order in
$\uv'$ and where $\gamma_{ij}^S= (\gamma_{ij} + \gamma_{ji} +
\gamma_{ik} \gamma_{jk})/2$. Since distortions at the boundary are
constrained to satisfy Eq.\ (\ref{boundary_1}), $u'_i ( \xv_B )$
is zero for all points $\xv_B$ on the boundary. It is often useful
to consider periodic boundary conditions in which $\uv'( \xv)$ has
the same value (possibly not zero) on opposite sides of the
parallelepiped. This condition implies
\begin{equation}
\int d^d x \partial_j u'_i ( \xv ) = \oint d S_j u'_i = 0 .
\end{equation}

\subsection{Nonaffinity in $1d$}

\begin{figure}
\centerline{\includegraphics{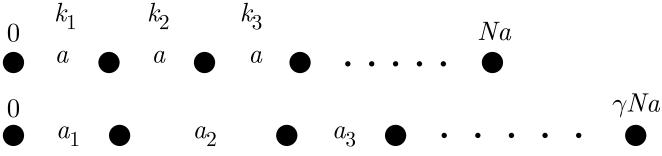}}
\caption{Schematic diagram of nonaffine distortion in a
one-dimensional lattice with random spring constants.  The top
figures shows the undistorted lattice of $N$ sites with random
spring constants $k_{\ell}$ and constant lattice spacing $a$. The
bottom figures shows that stretched lattice with length $\gamma
Na$ and random lattice spacings $a_{\ell} = \gamma + u'_{\ell}-
u'_{\ell -1}$.} \label{fig:1d_nonaffine}
\end{figure}

To develop quantitative measures of nonaffinity, it is useful to
consider a simple one-dimensional model, which can be solved
exactly. We study a one-dimensional periodic lattice, depicted in
Fig.\ \ref{fig:1d_nonaffine} with sites labelled by $\ell = 0 ,
... , N$, whose equilibrium positions are $R_{\ell 0} = a \ell$,
where $a$ is the rest bond length. Harmonic springs with spring
constant $k_\ell\equiv k + \delta k_\ell$ connect sites $\ell$ and
$\ell-1$, where $k=(\sum_{\ell} k_{\ell})/N$ is the average spring
constant and $\sum_\ell \delta k_{\ell} = 0$. The lattice is
stretched from its equilibrium length $N a$ to a new length $L =
\gamma N a$. If all $k_\ell$'s were equal, the lattice would
undergo an affine distortion with $R_\ell = \gamma a \ell$. When
the $k_\ell$'s are random, sites undergo an additional nonaffine
displacement $u'_\ell$ so that $R_\ell = \gamma a \ell + u'_\ell$.
The energy is thus
\begin{equation}
\Ha = \frac{1}{2} \sum_{\ell = 1}^N k_\ell ( \gamma a + u'_\ell -
u'_{\ell -1} )^2 .
\end{equation}
In equilibrium, the force $F_\ell = -\partial \Ha/\partial
u'_{\ell}$ on each bond is zero.  The resulting equation for
$u'_\ell$ is
\begin{equation}
F_{\ell} = k_{\ell+1}( \gamma a + u'_{\ell +1} - u'_{\ell} ) -
k_\ell (\gamma a + u'_{\ell} - u'_{\ell - 1} ) = 0 ,
\label{f_ell1}
\end{equation}
which can be rewritten as
\begin{equation}
- \Delta_+  k_\ell \Delta_- u'_{\ell} = \gamma a \Delta_+ \delta
k_l , \label{f_ell2}
\end{equation}
where $\Delta_+$ and $\Delta_-$ are difference operators defined
via $\Delta_+ A_\ell = A_{\ell + 1 } - A_{\ell}$ and $\Delta_- =
A_\ell - A_{\ell - 1}$ for any function $A_\ell$. The Fourier
transforms of $\Delta_+$ and $\Delta_-$ are, respectively,
$\Delta_+ ( q ) = e^{iq} - 1$ and $\Delta_- ( q ) = 1 - e^{-iq}$.
Equations (\ref{f_ell1}) and (\ref{f_ell2}) must be supplemented
with boundary conditions.  We use periodic boundary conditions for
which $u'_N = u'_0$ or equivalently
\begin{equation}
\sum_{\ell = 1}^N \Delta_- u'_{\ell} = 0 . \label{BC_1D}
\end{equation}
The solution to Eq.\ (\ref{f_ell2}) can be written as the sum of a
solution,
\begin{equation}
u_\ell^I = -(\Delta_+ k_\ell \Delta_-)^{-1} \gamma a \Delta_+
\delta k_\ell = -\gamma a \Delta_-^{-1} \frac{\delta
k_\ell}{k_\ell} , \label{eq:1Dinhomog}
\end{equation}
to the inhomogeneous equation and a solution,
\begin{equation}
\Delta_+ k_\ell \Delta_- u_\ell^H = 0 ,
\end{equation}
to the homogeneous one. The latter solution is $u_\ell^H =
\Delta_-^{-1} C/k_\ell$ where $C$ is an as yet undetermined
constant.   Adding the two solutions we obtain
\begin{equation}
u'_\ell =\Delta_-^{-1}\left( - \gamma a \frac{\delta
k_\ell}{k_\ell} + \frac{C}{k_\ell} \right) ,
\end{equation}
which implies $\Delta_- u'_\ell = (-\gamma a \delta k_\ell +
C)/k_\ell$.  The boundary condition of Eq.\ (\ref{BC_1D})
determines $C$, and the final solution for $u'_\ell$ is
\begin{align}
u'_{\ell} & = -\gamma a \Delta_-^{-1} \frac{1}{k_\ell}\left[\delta
k_\ell -
(\sum k_\ell^{-1} )^{-1} \sum k_\ell^{-1} \delta k_\ell \right] \nonumber \\
& =  \gamma a \Delta_-^{-1} \left( \frac{k_\ell^{-1}}{N^{-1} \sum
k_\ell^{-1}} - 1 \right) \equiv - \gamma a \Delta_-^{-1} S_\ell .
\label{u_1D}
\end{align}
The quantity
\begin{equation}
S_\ell = 1-\frac{(1+ p_\ell)^{-1}}{N^{-1} \sum_1^{\infty}(1+
p_{\ell})^{-1}}
\end{equation}
depends only on the ratio $p_{\ell} = \delta k_{\ell}/k$.

Equation (\ref{u_1D}) is the complete solution for $u'_\ell$ for
an arbitrary set of spring constants $k_\ell$.  In our model,
these spring constants are taken to be random variables, and
information about the nonaffinity is best represented by
correlation functions of the nonaffine displacement, averaged over
the ensemble of random $k_\ell$'s.  The simplest of these is the
two-point function $G(\ell - \ell') = \langle u'_{\ell'} u'_\ell
\rangle$, where $\langle \,\, \rangle$ represents an average over
$k_\ell$. $G(\ell - \ell')$ is easily calculated from Eq.\
(\ref{u_1D}); its Fourier transform is
\begin{equation}
G(q) = (\gamma a)^2 \frac{ \Delta^S ( q)}{2(1 - \cos q)} ,
\label{G(q)}
\end{equation}
where $\Delta^S ( q)$ is the Fourier transform of $\Delta^S (
\ell' - \ell ) = \langle S_{\ell'} S_\ell \rangle$.

There are several important observations that follow from the
expression Eq.\ (\ref{G(q)}) and that generalize to higher
dimensions.
\\
\newline
\noindent $\bullet$ $\Delta^S ( \ell', \ell)$ depends only on the
ratios $\delta k_\ell/ k$ and $\delta k_{\ell'}/k$, and it
increases with increasing width of the distribution of $\delta
k_\ell$. To lowest order in averages in $\delta k_\ell$,
\begin{eqnarray}
\Delta^S ( \ell', \ell) & = & k^{-2} [\Delta^k ( \ell' , \ell) -
N^{-1} \sum_{\ell_1} \Delta^k (\ell, \ell_1)] \nonumber
\\
&\rightarrow &k^{-2} \Delta^k ( \delta_{\ell, \ell'} - N^{-1} ) ,
\end{eqnarray}
where $\Delta^k ( \ell' , \ell ) = \langle \delta k_{\ell'} \delta
k_{\ell} \rangle$. The final form applies to uncorrelated
distributions in which spring constants on different bonds are
independent and $\langle \delta k_{\ell} \delta k_{\ell'} \rangle
= \Delta^k \delta_{\ell, \ell'}$. As the width of the distribution
increases, higher moments in $\delta k_{\ell}$ become important in
$\Delta^S ( \ell', \ell )$. If we assume that the only
nonvanishing fourth order moments are of the form
$\Delta^{(k,4)}(\ell', \ell) = \langle (\delta k_{\ell'})^2
(\delta k_{\ell} )^2 \rangle$, then the fourth-order contributions
to $\Delta^S$ are
\begin{align}
& k^4 \Delta^{(S,4)} ( \ell', \ell)  \nonumber \\
& =\left(1-\frac{4}{N}+\frac{6}{N^3}\right)\left[\Delta^{(k,4)}
(\ell',
\ell)-\frac{1}{N} \sum_{\ell_1}\Delta^{(k,4)}(\ell', \ell_1)\right]\nonumber \\
& + 2\left(1-\frac{3}{N^2}\right) \Delta^{(k,4)} (0)\left(
\delta_{\ell',\ell} - \frac{1}{N} \right) \nonumber \\
&  -\frac{1}{N}\left( 2 - \frac{3}{N}\right)
\sum_{\ell_1}\Delta^{(k,4)}(\ell,
\ell_1)\left(\delta_{\ell,\ell'}- \frac{1}{N}\right) .
\end{align}
For uncorrelated distributions, $\Delta^{(k,4)}(\ell, \ell') =
\Delta^{(k,4)} \delta_{\ell, \ell'} + (\Delta^k)^2 ( 1 -
\delta_{\ell,\ell'})$.  Thus, for uncorrelated distributions in
the limit $N \rightarrow \infty$,
\begin{eqnarray}
\Delta^{S} ( q) = \left(\frac{\Delta^k}{k^2} + \frac{3
\Delta^{(k,4)}-(\Delta^k)^2}{k^4}\right)(1- \delta_{q,0})
\label{quartic_Delta}
\end{eqnarray}
to fourth order in $\delta k_\ell$. Note that the constraint
$\sum_{\ell} \delta k_{\ell} = 0$ requires $\sum_{\ell}\Delta^S
(\ell, \ell') = \sum_{\ell'} \Delta^S ( \ell, \ell') =0$ and thus
that $\Delta^S (q=0) = 0$.  This condition is imposed by the
factor $1- \delta_{q,0}$ in Eq.\ (\ref{quartic_Delta}), which
implies that $\lim_{q\rightarrow 0} \Delta^S(q) \neq \Delta^S(q=0)
= 0$, i.e.,  $\Delta^S(q)$ does not approach zero as some power of
$q$ as $q\rightarrow 0$.

It is easy to verify that Eq.\ (\ref{quartic_Delta}) is exactly
the same result that would have been obtained using only the
solution [Eq.\ (\ref{eq:1Dinhomog})] to the {\em inhomogeneous}
equation for $u'_\ell$ with $\delta k_\ell/k_\ell$ replaced by
$\delta k_\ell/k_\ell - N^{-1} \sum_\ell \delta k_\ell/k_\ell$.
Thus, to obtain the solution for $G(q)$ to leading order $1/N$, we
can ignore the boundary condition, Eq.\ (\ref{BC_1D}), and use the
solution to the inhomogeneous equation with the constraint that
$q^2G(q)$ be zero at $q=0$.  This observation will considerably
simplify our analysis of the more complicated higher-dimensional
problem.
\\
\newline
$\bullet$ If correlations in $\delta k_\ell$ are of finite range,
then $\Delta^S (q) $ has a well defined $q\rightarrow 0$ limit. In
this limit,
\begin{equation}
G(q) = (\gamma a)^2 \frac{\Delta^S ( 0 )}{q^2} \approx
\frac{(\gamma a)^2}{q^2 k^2} \Delta^k ( q=0) ,
\end{equation}
where $\Delta^A(0) \equiv \lim_{q\rightarrow 0} \Delta^A(q)$ for
$A=S,k$. Thus, there is a $q^{-2}$ divergence in $G(q)$, and the
spatial correlation function $\mathcal{G} ( \ell' , \ell ) =
\langle (u_\ell' - u_\ell )^2 \rangle$ diverges linearly in
separation
\begin{equation}
\mathcal{G}(\ell, \ell') \sim (\gamma a)^2 \Delta^S ( 0 ) |\ell' -
\ell | \sim (\gamma a)^2 \frac{\Delta^k ( 0 )}{k^2} |\ell' - \ell
| .
\end{equation}
\\
\newline
$\bullet$ If correlations in $S_{\ell}$ extend out to a distance
$\xi$, then $\Delta^S ( q \xi)$ becomes a function  of $q \xi$.
Long-range correlations in $S_{\ell}$ will lead to long range
correlations in $G(q) \sim q^{-2} \Delta^S ( q \xi)$, and
$\mathcal{G} (\ell, 0)$ will grow more rapidly than $\ell$ for $1
\ll \ell \ll \xi$.  It is possible that this is the correlation
length that diverges at the jamming point $J$ in granular media
\cite{WyartWit2005,SilbertNag2005}. We will discuss this point
further in Sec.\ \ref{sec:long-range-corr}.

\subsection{Nonaffinity for $d>1$}

The nonaffinity correlation function,
\begin{equation}
G_{ij} ( \xv , \xv') = \langle u'_i ( \xv ) u'_j ( \xv')\rangle ,
\end{equation}
for $d>1$ has a form very similar to that for $d=1$, except that
it has more complex tensor indices. We will be primarily
interested in the scalar part of this function, obtained by
tracing over the indices $i$ and $j$.  The Fourier transform of
this function scales as
\begin{equation}
G( \qv ) \equiv G_{ii} ( \qv ) \sim \frac{\gamma^2}{q^2} \Delta^S
( \qv) \sim \frac{\gamma^2}{q^2} \frac{\Delta^K ( \qv )}{K^2} ,
\label{eq:G_23d}
\end{equation}
where $\gamma$ represents the appropriate components of the
applied strain and $\Delta^S ( \xv, \xv')$ is in general a
nonlinear function of the ratio of the fluctuating components
$\delta K_{ijkl} ( \xv )$ of the elastic-modulus tensor to its
uniform components $K_{ijkl}$.  To lowest order in the variance,
$\Delta^S \sim \Delta^K /K^2$ where $\Delta^K$ represents
components of the variance of the elastic-modulus tensor and $K$
components of its average. Thus, the nonaffinity correlation
function in coordinate space is proportional to $|\xv|^{-(d-2)}$
in dimension $d$, or
\begin{subequations}
\label{eq:calG_23d}
\begin{eqnarray}
\mathcal{G} ( \xv ) & = & \langle(\uv' ( \xv ) - \uv' (0))^2 \rangle \\
& \sim & A \ln (|\xv|/B)  \qquad d=2 \label{eq:calG_2d} \\
& \sim & C - D| \xv|^{-1}  \qquad d=3 ,
\end{eqnarray}
\end{subequations}
where
\begin{subequations}
\begin{eqnarray}
A& =&\frac{1}{\pi}\gamma^2 \Delta^S(0) \approx \frac{1}{\pi}
\gamma^2 \frac{\Delta^K (0)}{K^2} \\
B & = & (\alpha \Lambda)^{-1} \\
C &= &\gamma^2 \Delta^S(0) \frac{\Lambda}{\pi^2} \\
D &= & \frac{1}{\pi} \gamma^2 \Delta^S(0) ,
\end{eqnarray}
\end{subequations}
where $\Lambda = 2 \pi/a$ is the upper momentum cutoff for a
spherical Brillouin zone with $a$ the short distance cutoff and
$\alpha = 0.8905$ is evaluated in App.\ \ref{App:C}. The length
$B$ depends on the spatial form and range $\xi$ of local
elastic-modulus correlations.  We will derive explicit forms for
it shortly. In our numerical simulations, we allow the bond spring
constant $k_b$ to be a random variable with variance $\Delta^k =
\langle (\delta k_b )^2 \rangle$. Variations in $k_b$ in general
induce changes in all of the components of $\delta K_{ijkl}$, and
$\Delta^S$ is an average of a function of $\delta k_b/k$ where $k$
is the average of $k_b$.

In general $\gfunc$ also has anisotropic contributions whose
angular average is zero.  We will not consider these contributions
in detail, but we do evaluate them analytically in App.\
\ref{App:C}.

When a sample is subjected to a distortion via stresses at its
boundaries, the strains can be expressed in terms of an affine
strain and deviations from it.  Using the expressions in Eq.\
(\ref{eq:u_displace}) for these strains, we obtain the energy
\begin{eqnarray}
\delta \Ha & = & \frac{1}{2} \int \{K_{ijkl} \partial_j u'_i
\partial_l u'_k \nonumber \\
& & + [\delta K_{ijkl} ( \xv ) + \delta_{ik}\sigmat_{jl}(\xv)]
\partial_j u'_i \partial_l u'_k
\nonumber \\
& & + 2 \delta K_{ijkl} (\xv ) \gamma_{kl} \partial_j u'_i \}
\end{eqnarray}
to lowest order in $\gamma_{ij}$. Minimizing $\delta \Ha$ with
respect to $\uv'$, we obtain
\begin{equation}
- \partial_j [K_{ijkl} + \delta K_{ijkl}( \xv ) + \delta_{ik}
\sigmat_{jl} ( \xv )] \partial_l u'_k =
\partial_j \delta K_{ijkl} ( \xv ) \gamma_{kl} .
\label{eq:u'}
\end{equation}
This equation shows that the random part of the elastic-modulus
tensor times the affine strain acts as a source that drives
nonaffine distortions.  The random stress, which is transverse,
does not drive nonaffinity; it is the continuum limit of the
random force. The operator $-\partial_j K_{ijkl}^T (\xv
)\partial_l \delta(\xv - \xv')\equiv \chi_{ik}^{-1} ( \xv, \xv')$,
where $K_{ijkl}^T (\xv )= K_{ijkl} + \delta K_{ijkl}( \xv ) +
\delta_{ik} \sigmat_{jl} ( \xv )$ is the continuum limit of the
dynamical matrix or Hessian discussed in Refs.\
\cite{LemaitreMal2005} and \cite{TanguyBar2002}. The matrix
$\chi_{ij} ( \xv, \xv')$ is the response $\delta u_i ( \xv
)/\delta f_i ( \xv')$ of the displacement to an external force.
The formal solution to Eq.\ (\ref{eq:u'}) for $u'_i ( \xv)$ in
terms of $\delta K_{ijkl} ( \xv )$ and $\sigmat_{ij} ( \xv )$ is
trivially obtained by operating on both sides with $\chi_{pi} (
\xv, \xv')$:
\begin{equation}
u'_i ( \xv ) = \int d^d x' \chi_{ip} ( \xv - \xv')\partial'_j
\delta K_{pjkl} (\xv' ) \gamma_{kl},
\label{eq:sol_u'}
\end{equation}
The random component of the elastic modulus appears both
explicitly and in a hidden form in $\chi_{ip}$ in this equation.

Equation (\ref{eq:sol_u'}) is the solution to the inhomogeneous
equation, Eq.\ (\ref{eq:u'}).  Solutions to the homogeneous
equation should be added to Eq.\ (\ref{eq:sol_u'}) to ensure that
the boundary condition $\uv'( \xv_B)=0$ for points $\xv_B$ on the
sample boundary is met.  As in the $1D$ case, however, the
contribution from the homogeneous solution vanishes in the
infinite volume limit and can be ignored.

To lowest order in the randomness, we replace $\chi_{ip}$ in Eq.\
(\ref{eq:sol_u'}) with its nonrandom counterpart, $\chi_{ip}^0 (
\xv - \xv')$, the harmonic elastic response function $\delta u_i
(\xv)/\delta f_j ( \xv')$ of a spatially uniform system with
elastic-modulus tensor $K_{ijkl}$ to an external force $f_j (
\xv')$. Thus, to lowest order in $\gamma_{ij}$,
\begin{equation}
G_{ii'}( \qv ) = \chi_{ip}^0(\qv) \chi_{i'p'}^0(-\qv ) q_j q_{j'}
\Delta_{pjkl;p'j'k'l'}^K ( \qv) \gamma_{kl}\gamma_{k'l'} ,
\label{eq:full_G}
\end{equation}
where
\begin{equation}
\Delta_{ijkl;i'j'k'l'}^K (\xv , \xv')= \langle \delta K_{ijkl} (
\xv ) \delta K_{i'j'k'l'} ( \xv' )\rangle
\end{equation}
is the variance of the elastic-modulus tensor, which we simply
call the {\bf modulus correlator}. Equation (\ref{eq:full_G})
contains all relevant information about nonaffine correlations to
lowest order in the imposed strain.  It applies to any system with
random elastic moduli and stresses regardless of the symmetry of
its average macroscopic state.

Our primary interest is in systems whose elastic-modulus tensor is
macroscopically isotropic. In these systems, which include
two-dimensional hexagonal lattices, $K_{ijkl} = \lambda
\delta_{ij}\delta_{kl} + \mu (\delta_{ik} \delta_{jl} +
\delta_{il} \delta_{jk})$ is characterized by only two elastic
moduli, the shear modulus $\mu$ and the bulk modulus $B= \lambda
+(2 \mu/d)$, where $d$ is the dimension of the reference space.
The Fourier transform of $\chi_{ij}^0 ( \xv , \xv')$ in an
isotropic system is
\begin{equation}
\chi_{ij}^0 ( \qv ) = \frac{1}{(\lambda+ 2 \mu) q^2} {\hat q}_i
{\hat q}_j + \frac{1}{\mu q^2} (\delta_{ij} -{\hat q}_i {\hat q}_j
) . \label{eq:chi0}
\end{equation}
The modulus correlator is an eighth-rank tensor. At $\qv=0$, it
has eight independent components in an isotropic medium (See App.
\ref{App:A}) and more in media with lower symmetry, including
hexagonal symmetry. As discussed above, however, all components of
$\delta K_{ijkl}$ are proportional to $\delta k_b$.

We show in App. \ref{App:B} that $G(\qv)$ has the general form
\begin{equation}
G(\qv) = \frac{\gamma_{xy}^2}{\mu^2 q^2} (\Delta_A + \Delta_B
{\hat q}_{\perp}^2 - \Delta_C {\hat q}_x^2 {\hat q}_y^2 ) ,
\label{eq:G_aniso}
\end{equation}
where ${\hat q}_i = q_i/q$, $\hatq_{\perp}^2 = {\hat q}_x^2 +
{\hat q}_y^2$ and $\Delta_A$, $\Delta_B$ and $\Delta_C$ are linear
combinations of the independent components of
$\Delta^K_{ijkl;i'j'k'l'}$ times a function of $\lambda/\mu$.
Thus, in general $\gfunc$ will have anisotropic parts that depend
on the direction of $\xv$ in addition to an isotropic part that
depends only on the magnitude of $\xv$. In App. \ref{App:C}, we
derive expressions for the full form of $\gfunc$.  Here we discuss
only the isotropic part, which has the from of Eq.\
(\ref{eq:calG_23d}) with
\begin{align}
A& = \frac{\gamma_{xy}^2}{\pi \mu^2}[\Delta_A + \Delta_B -
\frac{1}{8}\Delta_C] \\
B& = (\alpha \Lambda)^{-1} \\
C& =\frac{\gamma_{xy}^2}{\pi^2 \mu^2}[\Delta_A + \frac{2}{3}
\Delta_B - \frac{1}{15}\Delta_C] \Lambda \\
D& =\frac{\gamma_{xy}^2}{2 \pi \mu^2}[\Delta_A + \frac{2}{3}
\Delta_B - \frac{1}{15}\Delta_C] .
\end{align}
In two dimensions, the anisotropic term is proportional to $\cos 4
\psi$ where $\psi$ is the angle that $\xv$ makes with the
$x$-axis.  In the limit of large $|\xv|$, the coefficient of $\cos
4 \psi$ is a constant.  In three dimensions, the anisotropic terms
are more complicated.  In both two and three dimensions, however,
the average of the anisotropic terms over angles are zero.

\subsection{\label{sec:other_nonaffine}Other Measures of Nonaffinity}

The nonaffinity correlation function $G_{ij}$ (and its cousin
$\cal G$) is not the only measures of nonafinity, though other
measures can usually be represented in terms of it.  Perhaps the
simplest measure of nonaffinity is simply the mean-square
fluctuation in the local value of of $\uv' ( \xv )$, which is the
equal-argument limit of the trace of $G_{ij} ( \xv' , \xv)$:
\begin{equation}
\langle [u' ( \xv )]^2 \rangle = G_{ii} ( \xv , \xv ) .
\end{equation}
This measure was used in Ref.\ \cite{LangerLiu1997} to measure
nonaffinity in models for foams. In three dimensions, it is a
number that depends on the cutoff, $a^{-1}$: $\langle [u' ( \xv
)]^2 \rangle \sim \gamma^2 (\Delta^K/K^2) a^{-1}$;  in two
dimensions, it diverges logarithmically with the size of the
sample $L$: $\langle [u' ( \xv )]^2 \rangle \sim \gamma^2
(\Delta^K/K^2)\ln (L/a)$.

\begin{figure}
\centerline{\includegraphics{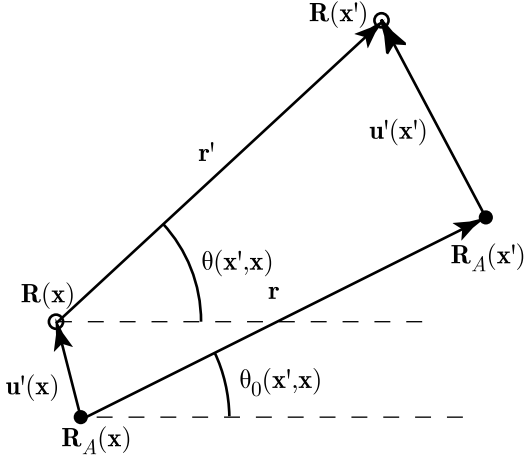}}
\caption{Graphical representation of the angular measure of nonaffinity
used in Ref.\ \cite{HeadMac2003c}.  Under affine distortions,
points $x_i$ transform to points $R_{Ai} ( \xv ) = x_i +
\gamma_{ij} x_j$, and the vector $\rv = \Rv_A ( \xv' ) - \Rv_A(
\xv )$ makes and angle $\theta_0 ( \xv' , \xv )$ with the
$x$-axis. Under nonaffine distortions, points $x_i$ transform to
$R_i(\xv) = R_{Ai} ( \xv) + u'_i ( \xv)$, and the vector $\rv' =
\Rv( \xv') - \Rv( \xv)$ makes an angle $\theta ( \xv' , \xv )$
with the $x$-axis.} \label{fig:nonaffine_angle}
\end{figure}

References \cite{HeadMac2003c,HeadMac2003a,HeadMac2003b}, which
investigate a two-dimensional model of crosslinked semi-flexible
rods designed to describe crosslinked networks of actin and other
semi-flexible biopolymers, introduce [Fig.\
\ref{fig:nonaffine_angle}] a measure based on comparing the angle
$\theta \equiv \theta( \xv' , \xv)$ that the vector connecting two
sites originally at $\xv$ and $\xv'$ makes with some fixed axis
after nonaffine distortion under shear to the angle $\theta_0
\equiv \theta_0( \xv') - \theta_0 ( \xv)$ that that vector would
make if the points were affinely distorted:
\begin{equation}
\mathcal{G}_{\theta}( \xv' - \xv) = \langle [\theta( \xv', \xv) -
\theta_0(\xv', \xv)]^2 \rangle . \label{eq:G_theta}
\end{equation}
Under affine distortion, the vector connecting points $\xv'$ and
$\xv$ is $r_i = x'_i - x_i + \gamma_{ij}(x'_j - x_j)$; under
nonaffine distortion, the separation is $\rv' = \rv + \uv' ( \xv')
- \uv'( \xv )$.  In two dimensions,
\begin{equation}
\rv \times \rv' = r r' \sin(\theta - \theta_0) \ev_z = \rv \times
[ \uv' ( \xv') - \uv' ( \xv)] ,
\end{equation}
where $\ev_z$ is the unit vector along the $z$ direction
perpendicular to the two-dimensional plane and $r = |\rv|$.  If
both $\gamma$ and $|\uv'(\xv') - \uv'(\xv)|/|\xv'-\xv|$ are small,
\begin{equation}
\theta ( \xv' , \xv ) - \theta_0 ( \xv' , \xv ) \approx
\frac{\ev_z \cdot [(\xv' - \xv ) \times (\uv'( \xv' ) - \uv' ( \xv
))]}{|\xv' - \xv |^2} ,
\end{equation}
and
\begin{equation}
\mathcal{G}_\theta (\xv) = \epsilon_{ij} \epsilon_{kl} \frac{x_i
x_j}{|\xv|^4}\mathcal{G}_{kl}(\xv) \sim \frac{1}{|\xv |^2}
\mathcal{G}(\xv) \label{eq:G_theta_G}
\end{equation}
where $\epsilon_{ij} = \epsilon_{zij}$ is the two-dimensional
antisymmetric symbol, and $\mathcal{G}_{ij} ( \xv ) = \langle
[u'_i ( \xv) - u'_i ( 0 ) ][u'_j(\xv) - u'_j ( 0 ) ]\rangle$.

\subsection{\label{sec:ran_stress_1}Generation of Random Stresses}

As we have discussed, a system of particles in mechanical
equilibrium can be characterized by random elastic moduli and a
random local stress tensor with only transverse components. To
better understand random stresses, it is useful to consider a
model in which random stress is introduced in a material that is
initially stress free. We begin with a system with a local
elastic-modulus tensor $K_{ijkl} ( \xv)$ that can in general be
random but with $\sigmat_{ij} ( \xv ) = 0$, and to this we add a
local random stress $\sigmab_{ij}(\xv)$ with zero mean that
couples to the rotationally invariant nonlinear strain and that
has longitudinal components so that its variance in an isotropic
system is
\begin{equation}
\Delta^{\sigmab}_{ijkl} = \Delta_1^{\sigmab} \delta_{ij}
\delta_{kl} + \Delta_2^{\sigmab}(\delta_{ik}\delta_{jl} +
\delta_{il}+ \delta_{jk} ) .
\end{equation}
For simplicity, we assume that the spatial average of
$\sigmab_{ij} ( \xv)$ is zero. A random stress of this sort can be
generated in a lattice model by making the rest bond length
$R_{bR}$ a random variable in a system in which initially the rest
and equilibrium bond lengths are equal. In the continuum limit,
our elastic energy is thus
\begin{equation}
\delta \Ha = \int [\textstyle{\frac{1}{2}}K_{ijkl}^\sigma (\xv)
u_{ij} ( \xv ) u_{kl} ( \xv ) + \sigmab_{ij} ( \xv ) u_{ij} ( \xv)
] ,
\label{eq:sigma_full}
\end{equation}
where the superscipt $\sigma$ on $K_{ijkl}^{\sigma}$ indicates
that this is an elastic modulus prior to relaxation in the
presence of $\sigma_{ij}$.

Sites that were in equilibrium at positions $\xv$ in the original
reference space in the absence of $\sigmab_{ij}$ are no longer so
in its presence.  These sites will undergo displacements to new
equilibrium sites $\xv' \equiv \Rv_0 ( \xv)= \xv + \uv_0 ( \xv )$,
which define a new reference space. Positions $\Rv( \xv ) $ in the
target space can be expressed as displacements relative to the new
reference space: $\Rv(\xv' ) = \xv' + \uv' ( \xv' )$.  Then,
strains relative to the original reference space can be expressed
as the sum of a strain relative to the new reference space and one
describing the distortion of the original references space to the
new one:
\begin{eqnarray}
u_{ij} ( \xv ) & = &\frac{1}{2} \left(\frac{\partial R_k (
\xv)}{\partial x_i} \frac{\partial R_k ( \xv ) }{\partial
x_j} - \delta_{ij} \right)\nonumber \\
& = &u_{ij}^0 (\xv) + \Lambda_{0ik}^T ( \xv' ) u'_{kl} ( \xv' )
\Lambda_{0lj}( \xv')  , \label{eq:strain'}
\end{eqnarray}
where
\begin{eqnarray}
\Lambda_{0ij}( \xv ) & = &\frac{\partial R_{0i} ( \xv )}{\partial
x_j} = \delta_{ij} + \partial_j u_{0i}, \\
u_{ij}^0 & = & \frac{1}{2} \left(\Lambda_{0ki} \Lambda_{0kj} -
\delta_{ij} \right) ,
\end{eqnarray}
and
\begin{equation}
u'_{ij} ( \xv' ) = (\partial'_i u'_j + \partial'_j u'_i +
\partial'_i u'_k \partial'_j u'_k )/2 ,
\end{equation}
where $\partial'_i \equiv \partial/\partial x'_i$.  Using Eq.\
(\ref{eq:strain'}) in Eq.\ (\ref{eq:sigma_full}), we obtain
$\delta \Ha [\uv] =\delta \Ha[\uv_0] + \delta \Ha'[\uv']$, where
\begin{equation}
\delta \Ha'  =  \frac{1}{2} \int d^d x'[K_{ijkl}u'_{ij} u'_{kl} +
\sigmat_{jl}u'_{ij}]
\end{equation}
with
\begin{equation}
K_{ijkl} ( \xv' ) = (\det \Lambda_0)^{-1}\Lambda_{0ia}
\Lambda_{0jb}K_{abcd}^\sigma \Lambda_{0ck}^T \Lambda_{0dl}^T
\end{equation}
and
\begin{equation}
\sigmat_{ij} (\xv') = (\det \Lambda_0)^{-1}\Lambda_{0ia}(
K_{abcd}^\sigma u_{0cd} +\sigmab_{ab} )\Lambda_{0bj}^T ,
\end{equation}
where we have not displayed explicitly the dependence of
$\Lambda_{0ij}$ on $\xv'$. The displacement field $\uv_0 ( \xv)$
is determined by the condition that the force density at each
point in the new reference state be zero, i.e., so that
$\partial'_j \sigmat_{ij} ( \xv ' ) = 0$. To linear order in
displacement and $\sigmab_{ij}$, this condition is
\begin{equation}
\partial_j ( K_{ijkl} u_{0kl} + \sigmab_{ij} ) = 0 ,
\end{equation}
where to this linearized order, we can ignore the difference
between $\xv$ and $\xv'$. For an initially isotropic medium, this
equation can be solved for $\uv_0$ to yield
\begin{equation}
u_{0i}(\qv) = \frac{1}{\mu q^2}\left(\delta_{ik} - \frac{\lambda+
\mu}{\lambda+2 \mu} \frac{q_i q_k}{q^2}\right) i q_l \sigmab_{kl}
.
\label{eq:u_sigma}
\end{equation}
To lowest order in $\uv_0$, the elastic moduli and stress tensors
in the new reference state are
\begin{eqnarray}
\sigmat_{ij} ( \qv ) & = & \delta_{ik}^T \delta_{jl}^T
\sigmab_{kl} - \frac{\lambda}{\lambda + 2 \mu}
\delta_{ij}^T {\hat q}_k {\hat q}_l \sigmab_{kl} \\
\delta K_{ijkl} & = & 2 \lambda (\delta_{ij} v_{kl} +
\delta_{kl} v_{ij}) \\
& & + 2 \mu(\delta_{ik} v_{jl} + \delta_{jl} v_{ik} + \delta_{il}
v_{jk} + \delta_{jk} v_{il} ) , \nonumber
\end{eqnarray}
where
\begin{eqnarray}
v_{ij} & = & \frac{1}{2}\left( \Lambda_{0ik} \Lambda_{0kj}^T -
\delta_{ij} \right) \nonumber \\
& = & (\partial_i u_{0j} + \partial_j u_{0i} +
\partial_k u_{0i} \partial_k u_{0j})/2 \nonumber
\\
& \approx & (\partial_i u_{0j} + \partial_j u_{0i} )/2
\label{eq:vij}
\end{eqnarray}
is the left Cauchy strain tensor relative to the original
reference state.

Note that $\sigmat_{ij} ( \xv')$ is transverse and random as it
should be.  The elastic-modulus tensor is a random variable via
its dependence on $\Lambda_{0ij} ( \xv )$. Thus, a random stress
added to an initially homogeneous elastic medium (with
$K_{ijkl}^0$ nonrandom and independent of $\xv$) produces both a
random transverse stress and a random elastic-modulus tensor in
the new relaxed reference frame. The statistical properties of
$K_{ijkl} ( \xv')$ are determined in this model entirely by those
of $\sigmab_{ij} ( \xv)$, and $\Delta^K \sim \Delta^{\overline
\sigma}$. In general, of course, the randomness in $K_{ijkl}(
\xv')$ arises both from randomness in the original
$K_{ijkl}^{\sigma} ( \xv )$ and $\sigmab_{ij} ( \xv )$.

The nonaffinity correlation function can be calculated exactly to
lowest order in $\Delta^{\sigmab}_1$ and $\Delta^{\sigmab}_2$ when
the initial reference state is homogeneous and nonrandom.  It has
exactly the same form as Eq.\ (\ref{eq:G_23d}) when expressed in
terms of $\Delta^K$. When expressed in terms of
$\Delta^{\sigmab}_1$ and $\Delta^{\sigmab}_2$, it has a similar
form, which in an isotopic elastic medium can be expressed as
\begin{equation}
G( \qv ) \sim \frac{\Delta^{\sigmab}_2 \gamma^2}{\mu^2 q^2} f(
{\hat \qv}, \lambda/\mu, \Delta^{\sigmab}_1/ \Delta^{\sigmab}_2) .
\end{equation}
Thus, $\mathcal{G}( \xv )$ has the same form in this model as Eq.\
(\ref{eq:calG_23d}).

\subsection{\label{sec:long-range-corr}Long-range Correlations in Elastic Moduli}

Long-range correlations in random elastic moduli can significantly
modify the behavior of $\mathcal{G} ( \xv )$. To illustrate this,
we consider a simple scaling form for $\Delta^K ( \qv)$ inspired
by critical phenomena:
\begin{align}
&\Delta^K ( \qv ) =  \xi^{\phi} g( q \xi ) \label{eq:limit_Del}\\
&\sim
    \begin{cases}
    \xi^{\phi} g_0 [1 + (q \xi)^s + ... ],& \text{for $q\xi \rightarrow
    0$},\\
    g_{\infty} q^{-\phi} [1 + b (q \xi)^{-t} + ... ],& \text{as $ q \xi
    \rightarrow \infty$} ,
    \end{cases}
\label{eq:ginfty}
\end{align}
where $\xi$ is a correlation length, $\phi$ is the dominant
critical exponent, and $s$ and $t$ are corrections to scaling
exponents.  It is possible in principle for each of the components
of $\Delta_{ijkl;i'j'k'l'}^K$ to be described by difference
scaling lengths $\xi$ and functions $g(u)$.  We will assume,
however, that $\xi$ and the functional form of $g$ is the same for
all components, but we will allow for the zero-argument value
$g_0$ to vary.  $G(\qv)$ is thus given by Eq.\ (\ref{eq:G_aniso})
with $\Delta_A$, $\Delta_B$, and $\Delta_C$ replaced by $\Delta_A
( \qv)$, $\Delta_B(\qv)$, and $\Delta_C ( \qv)$ with scaling forms
given by Eq.\ (\ref{eq:limit_Del}).  In this case, $\gfunc$ can be
written as $(\gamma_{xy}^2/\mu^2) \mathcal{F}(\xv)$ with
$\mathcal{F}(\xv) = \calF_A (\xv)+\calF_B(\xv) - \calF_C(\xv)$,
where
\begin{equation}
\calF_{\alpha} ( \xv ) = 2 \xi^{\phi} \int \frac{d^d q}{(2 \pi)^d}
f_\alpha ( \qv) g_\alpha ( q\xi) \frac{1}{q^2} (1- e^{i \qv \cdot
\xv}) ,
\label{eq:calF}
\end{equation}
with $f_A = 1$, $f_B(\qv ) = {\hat q}_{\perp}^2$, and $f_C(\qv) =
{\hat q}_x^2 {\hat q}_y^2$. There are two important observations
to make about the functions $\calF_{\alpha}$. First, for $q \xi
\ll 1$, $g(q \xi)$ can be replaced by its zero $q$ limit, $g_0$.
Thus, as long as $\xi$ is not infinite, the asymptotic behavior of
$\mathcal{G}(\xv)$ for $|\xv| \gg \xi$ is identical to those of
Eq.\ (\ref{eq:calG_23d}) but with amplitudes that increase as
$\xi^\phi$. Second, when $\xi \rightarrow \infty$, the $q^{(d-3-
\phi)}$ behavior of the integrand leads to modified power-law
behavior in $|\xv|$ for $a \ll x \ll \xi$, where $a = 2 \pi /
\Lambda$ is the short distance cutoff, depending on dimension.

In two dimensions, which is the focus of most of our simulations,
the isotropic part of $\calF$ is
\begin{equation}
{\cal F}_I (\xv ) = \frac{\xi^{\phi}}{\pi} \int_0^{\Lambda}
\frac{dq}{q} g( q \xi) [1 - J_0 ( q |\xv|)] ,
\label{eq:calF_I}
\end{equation}
where $g(y) = g_A(y) + g_B(y) - \frac{1}{8} g_C(y)$ and $J_0(y)$
is the zeroth order Bessel function. In the limit $|\xv| \gg \xi$,
\begin{equation}
\calF_I(\xv) \sim \frac{1}{\pi} g_0 \xi^{\phi} \ln
\frac{\beta(\Lambda \xi,\phi)}{\xi} |\xv| ,
\label{eq:calF_asy}
\end{equation}
where $\beta(\Lambda \xi, \phi)$ is evaluated in Appendix
\ref{App:C}. The behavior of $\calF_I (\xv)$ when $\Lambda^{-1}
\ll |\xv| \ll \xi$ depends on the value of $\phi$
\begin{equation}
\calF_I(\xv) \sim
    \begin{cases}
    \frac{1}{\pi}g_{\infty} |\xv|^\phi \mathcal{A}_2(\phi), & \text{if $\phi <2$} ,\\
    \frac{1}{4\pi}g_{\infty} |\xv|^2 \ln (\nu \xi/|\xv|), & \text{if $\phi
    =2$},\\
    \frac{1}{4\pi}g_{\infty} |\xv|^2 \xi^{\phi-2}
    \mathcal{C}_2(\phi), & \text{if $\phi >2$} .
    \end{cases}
\label{eq:2D_correlated}
\end{equation}
The quantities $\mathcal{A}_2(\phi)$, $\mathcal{C}_2(\phi)$, and
$\nu$ are evaluated in Appendix \ref{App:C}.

\begin{figure}
\centerline{\includegraphics{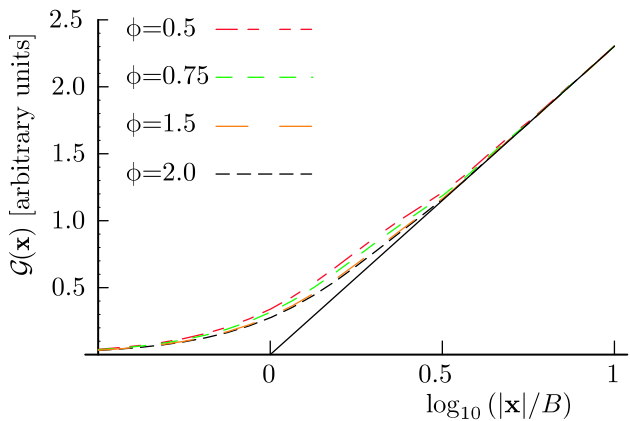}}
\caption{The function ${\cal G}(|\xv|)$ in two dimensions for
fixed $\gamma$ and $K$ for systems with long-range
correlations in $\Delta^K$ defined by Eqs.\ (\ref{eq:calF_I}) and
(\ref{eq:g(u)_1}) for different exponents $\phi$. The amplitudes
and correlation lengths $\xi$ for each $\phi$ are normalized so
that all curves have the same values of $B= \xi/\beta(\Lambda \xi,
\phi)$ and coefficients of $\ln |\xv|/B$ at large $|\xv|/B$. The
curve for $\phi=2$ corresponds to Lorentzian correlations [Eq.\
(\ref{eq:Lorentzian})]. [Color online] }
\label{fig:2d_long_range}
\end{figure}

The function $g(u)$ can have any form provided its large- and
small-$u$ limits are given by Eq.\ (\ref{eq:limit_Del}).  A useful
model form to consider, of course, is the simple Lorentzian for
which $\phi = 2$ and
\begin{equation}
g(u) = \frac{g_0}{1+ u^2} .
\label{eq:Lorentzian}
\end{equation}
For the purposes of illustration, in Fig.\ \ref{fig:2d_long_range}
we plot  ${\cal F}_i ( |\xv|)$ for a family of functions
parameterized by the exponent $\phi$:
\begin{equation}
g(u) = \frac{g_0(\phi)}{(1 + u^2)^{\phi/2}} .
\label{eq:g(u)_1}
\end{equation}
these curves clearly show the crossover from $|\xv|^{\phi}$
behavior for $\Lambda^{-1} \ll |\xv| \ll \xi$ to the
characteristic log behavior for $|\xv| \gg \xi$.  The correlation
length $\xi$ and the amplitude $g_0(\phi)$ were set so that the
large $\xv$ log behavior is the same for every $\phi$.  For this
family of crossover functions, the value of $|\xv|$ at which
${\cal F} ( |\xv|)$ crosses over from $|\xv|^{\phi}$ to
logarithmic behavior increases with decreasing $\phi$, and curves
with smaller $\phi$ systematically lie above those with large
$\phi$.

The limiting forms for ${\cal F}_I(|\xv|)$ in one and three
dimensions are given in App. \ref{App:C}.

\subsection{Rotational Correlations}

\begin{figure}
\centerline{\includegraphics[width=3in]{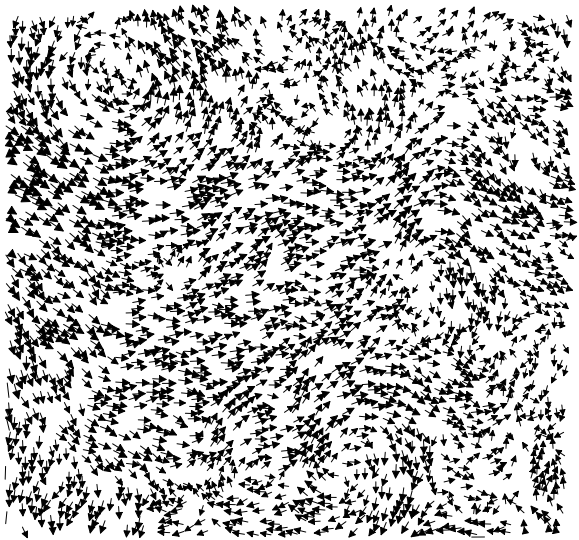}}
\caption{This figure shows the direction of nonaffine displacements resulting
from a shear in  the  $xy$-plane displayed as arrows on the
original reference lattice. Note the vortex-like patterns.}
\label{fig:rotations}
\end{figure}

The nonaffine displacements generated in random elastic media by
external strains contain rotational as well as irrotational
components as is evident from Fig.\ (\ref{fig:rotations}).  The
local nonaffine rotation angle is $\omega_k ( \xv ) = \frac{1}{2}
\epsilon_{ijk} \partial_j u'_k$, where $\epsilon_{ijk}$ is the
anti-symmetric Levi-Civita tensor, and rotational correlations are
measured by the correlation function $G_{\omega_i \omega_j} (\xv )
= \langle \omega_i ( \xv ) \omega_j ( 0 ) \rangle$.  In two
dimensions, there is only one angle $\omega (\xv) = \frac{1}{2}
\epsilon_{ri}
\partial_r u_i$, where $\epsilon_{ri} \equiv \epsilon_{zri}$.  The
Fourier transform of the correlation function $G_{\omega} =\langle
\omega ( \xv ) \omega ( 0 ) \rangle$ will then scale as $\gamma^2
\Delta^K/ \mu^2$, approaching a constant rather than diverging as
$\qv\rightarrow 0$.  We show in App.\ \ref{App:D} that
\begin{equation}
G_{\omega}( \qv ) = \frac{\gamma_{xy}^2}{\mu^2}
[\Delta_A^{\omega}(q) - \Delta^{\omega}_C(q) \hatq_x^2 \hatq_y^2]
\end{equation}
in two dimensions, where $\Delta_A^{\omega}(q)$ and
$\Delta_C^{\omega}$ are linear combinations of the independent
components of $\Delta^K_{ijkl;i'j'k'l'}$.  Thus, the rotation
correlation function contains direct information about
elastic-modulus correlations. If these correlations are short
range, and there is no $q$ dependence in either
$\Delta_A^{\omega}(q)$ or $\Delta_C^{\omega}$, the spatial
correlation function has an isotropic short-range part and an
anisotropic power-law part:
\begin{equation}
G_{\omega}( \xv ) =
\frac{\gamma_{xy}^2}{\mu^2}\left[\Delta_A^{\omega} \delta( \xv ) -
\Delta_C^{\omega} \left(16 \frac{x^2 y^2}{|\xv|^6} -
\frac{2}{|\xv|^2}\right) \right] .
\end{equation}
If there are long-range correlations in the elastic moduli with
the Lorentzian form of Eq.\ (\ref{eq:Lorentzian}), then
\begin{eqnarray}
G_{\omega}( \xv ) &= &\frac{\gamma_{xy}^2}{2 \pi \mu^2}\Bigl[
\left({\tilde \Delta}_A^{\omega} -
\frac{1}{8}\Delta_C^{\omega}\right) K_0(|\xv|/\xi)
\label{eq:vortex_corr}\\
& & + \frac{1}{8} \cos 4 \psi{\tilde \Delta}_C^{\omega}\left(-
\frac{48 \xi^4}{|\xv|^4} + \frac{4 \xi^2}{|\xv|^2} +
K_4(|\xv|/\xi)\right)\Bigr] , \nonumber
\end{eqnarray}
where $K_n( y)$ is the Bessel function of imaginary argument. The
$\cos 4 \psi$ behavior is for isotropic systems.  There will be
$\cos 6 \psi$ and higher order terms present in a hexagonal
lattice.  In Sec.\ \ref{Num_min}, we verify in numerical
simulations the exponential decay of the isotropic part of
$G_{\omega} ( |\xv|)$ in Model A with long-range correlations in
spring constants and the $|\xv|^{-2}$ behavior of the $\cos 4
\psi$ part of $G_{\omega} ( |\xv|)$ in Model C, which is
isotropic.

\section{\label{Num_min}Numerical Minimizations}

To further our understanding of nonaffinity in random lattices and
to verify our analytic predictions about them, we carried out a
series of numerical studies on models A-D  described in Sec.\
\ref{model_en}.  To carry out these studies, we began with an
initial lattice -- a periodic hexagonal or FCC lattice for models
A and B and a randomly tesallated lattice for models C and D.  We
assigned spring potentials $V_b(R_b)$ and rest bond lengths
$R_{bR}$ to each bond. To study nonaffinity, we subjected lattices
to shear and then numerically determined the minimum-energy
positions of all sites subject to periodic boundary condition.
The elastic energy of the lattice
was linearized about the affine shear state.
Interestingly, in this linearization
the value of the imposed shear, $\gamma$, factored out of our calculation,
so that $\bfu'(\bfx)$ was linear in $\gamma$ and thus $\gfunc$ was
automatically quadratic in $\gamma$.
We present below the procedures and results for each model.

\subsection{Model A}

In this model, the initial reference lattice is periodic, and the
rest bond length $R_{bR}$ is equal to the equilibrium lattice
parameter $R_{b0}$ for every bond, which we set equal to one. Each
bond is assigned an anharmonic potential
\begin{equation}
V_b ( R_b ) = \frac{1}{2}k_b(\delta R_b^2 + \delta R_b^4) ,
\label{eq:anharpot}
\end{equation}
where $\delta R_R = R_b - R_{b0} \equiv R_b - 1$ and the spring
constant $k_b$ is a random variable.  We chose $k_b = 1 + \delta
k_b$ where $\delta k_b$ is a random variable with zero mean lying
between $- \delta \kbar$ and $+ \delta \kbar$ with $\delta \kbar
<1$.

\subsubsection{Independent bonds on hexagonal and FCC lattice}

In the simplest versions of model A, the spring constant $k_b$ is
an independent random variable on each bond of a two-dimensional
hexagonal or a three-dimensional FCC lattice. We assign each bond
a random value of $\delta k_b$ chosen from a flat distribution
lying between $-\delta \kbar$ and $+\delta \kbar$.  Randomly
distributed spring constants give rise to random local elastic
moduli as defined by Eq.\ (\ref{eq:local_el_modulus}).
We verified that the distribution of the values of
the local shear modulus $K_{xyxy}$ on a hexagonal lattice
for different $\delta \kbar$ was well fit by a Gaussian
function with width linearly proportional to $\delta \kbar$.

\begin{figure}

\centerline{\includegraphics[width=2.5in]{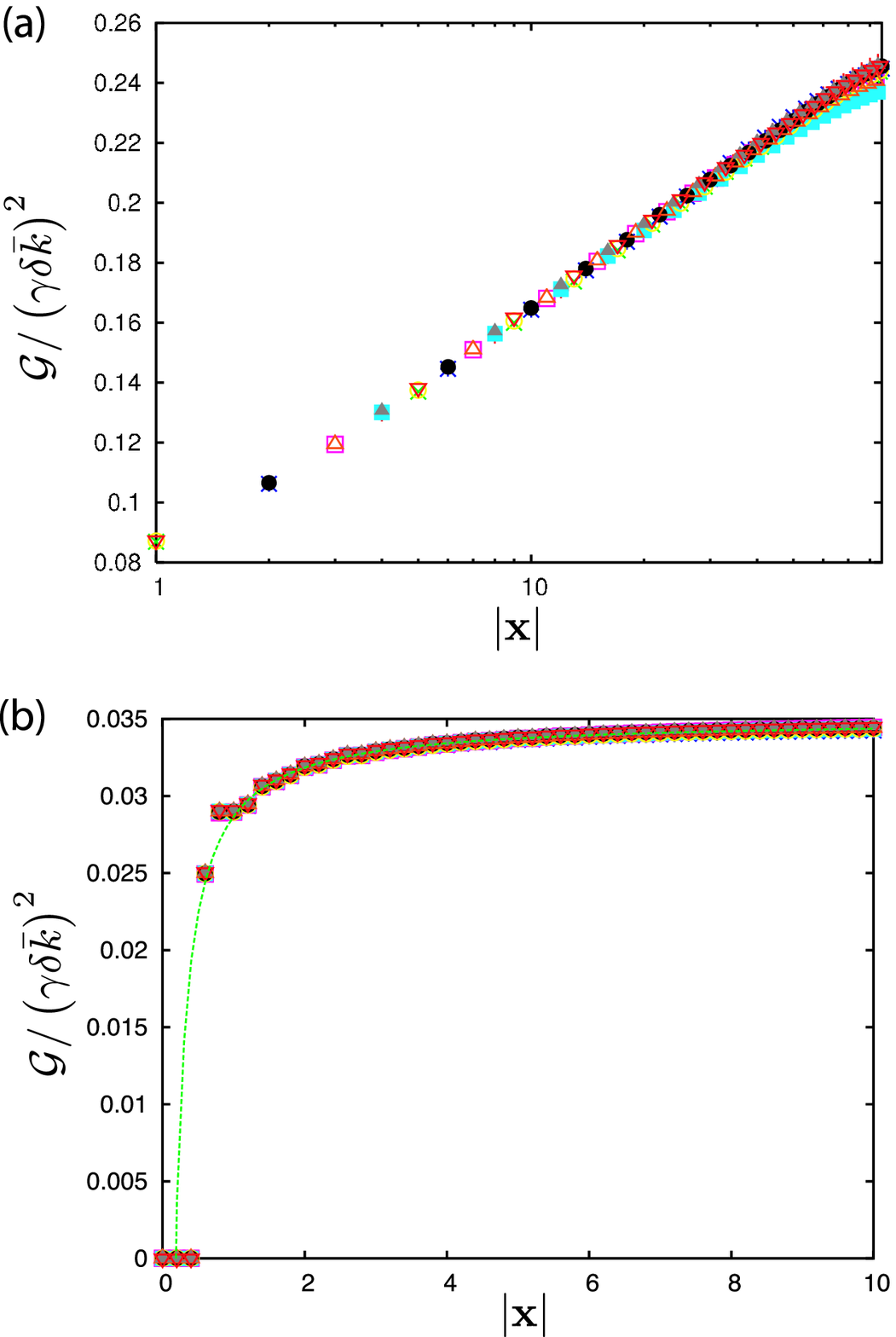}}

\caption{The nonaffinity correlation function
$\gfunc/(\gamma \delta \kbar)^2$ for sheared model-A
lattice: (a) $\gfunc/(\gamma \delta \kbar)^2$ versus
$\xsep$ for a $2$D hexagonal lattice sheared to
$0.1 \%$ for values $\delta \kbar = 0.01 n$
for $n = 1, ... , 10$ of the variation in local spring constant.
(b) $\gfunc/(\delta \kbar)^2$
versus $\xsep$ for a $3$D FCC lattice sheared to $0.1 \%$ for
several different values $\delta \kbar = 0.01 n$ for $n = 1, ... ,
10$. The data is fit to a function $C- D/\xsep$. All lengths are
in units of the lattice spacing. [Color online]
}
\label{fig:tom2c}
\end{figure}
\begin{figure}

\includegraphics[width=2.5in]{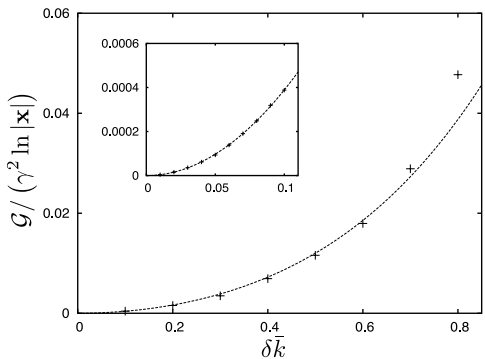}

\caption{(a)The prefactor A of the fits in plots of Fig.\
\ref{fig:tom2c}(a) as a function of $\delta \kbar$. The dotted
curve is a fit to  a quartic $g (\delta\kbar)^2 + h (\delta
\kbar)^4$ as suggested by Eq.\ (\ref{quartic_Delta}).   The inset
shows data up to $\delta \kbar = 0.1$ and a quadratic fit. }
\label{fig:hexamplitudes_1}
\end{figure}

The nonaffinity correlation function $\gfunc= \langle
\left|\bfu'(\xv) - \bfu'(0) \right|^2 \rangle$ [Eq.\
(\ref{eq:calG_23d})] measured on the numerically relaxed lattices
is shown in Fig.~\ref{fig:tom2c}(a). The averages were calculated by
summing the differences in deviation for every pair of nodes on
the lattice and binning according to the nodes' separation in the
undeformed (reference) state. Note that this process automatically
averages over angle, so it produces only the isotropic part of
$\gfunc$. The separation between nodes was taken as the least
distance between the nodes across any periodic boundaries. The
curves were well fit by the $A \ln (\xsep /B)$ dependence on
$\xsep$ predicted by Eq.\ (\ref{eq:calG_2d}). The excellent data
collapse achieved by plotting the rescaled function
$\gfunc/(\gamma \delta \kbar)^2$ demonstrates the quadratic
dependence of the amplitude $A$ on $\delta \kbar$.
Figure \ref{fig:hexamplitudes_1} shows the quadratic
plus quartic dependence of the amplitude $A$ on and $\delta \kbar$
at larger values. It is worth noting that while all correlation
functions were independently fit with a two-parameter function $A
\ln(\xsep/B)$, the optimal values of $B$ in all cases fell within
$10\%$ of one another.

Figure \ref{fig:tom2c}(b) displays $\gfunc/(\delta \kbar)^2$ on an
FCC lattice as a function of $\xsep$ for different $\delta \kbar$
for $\gamma = 0.1\%$ fit to the function $C - D/\xsep$ predicted
by Eq.\ (\ref{eq:calG_23d}). The data collapse verifies the
expected dependence of $C$ on $(\delta \kbar)^2$.

%
%




\subsubsection{Correlated random bonds on an hexagonal lattice}

\begin{figure}

\includegraphics[width=2.5in]{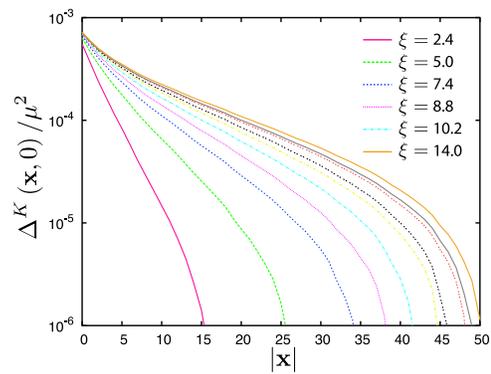}

\caption{The two point correlation function
$\Delta^K(\xv\!=\!0) = \langle \delta K_{xyxy}(\xv) \delta
K_{xyxy} (0) \rangle$ produced by random spring constants drawn
from an exponentially correlated random scalar field.  The
exponential decay is evident.  [Color online]}
\label{fig:corr_elas_const}
\end{figure}

As discussed in Sec.\ \ref{sec:long-range-corr}, random lattices
can exhibit long-range correlations, characterized by a
correlation length $\xi$, in local elastic moduli that can
significantly modify the behavior of nonaffinity correlation
functions at distances less than $\xi$.  To verify the prediction
of Sec.\ \ref{sec:ran_stress_1}, we numerically constructed
hexagonal lattices with long-range correlations in bond spring
constants. To do this, we set  $k_b = 1 + \delta k_b$ where
$\delta k_b$ was set equal to a small, randomly generated scalar
field with proper spatial correlations. This scalar field was
created by taking the reverse Fourier transform of the function
$\exp (i\phi_r)/\sqrt{q^2+\xi^{-2}}$, where $\xi$ is a variable
decay length and $\phi_r$ is a random complex phase. The scalar
field in these simulations was normalized to have constant mean
squared value and peak values of $\pm 0.1$, so that the variation
to the local spring constants was at most $10\%$. This method of
generation yields a clean exponential decay in the two-point
correlation function $\Delta^K(\xv,0)\equiv\langle \delta
K_{xyxy}(\xv) \delta K_{xyxy} (0) \rangle$ which persists for
separations up to three times the correlation length.
Figure~\ref{fig:corr_elas_const} shows the two-point correlation
function $\Delta^K(\xv,0)$ as a function of separation. The region
of exponential correlation was followed by a small region of
anti-correlation, which is not pictured. By construction, the
distributions of the local shear elastic modulus $K_{xyxy}$ were
essentially constant, independent of $\xi$; thus $\Delta^K(0,0)$
is equal for all curves in Fig.~\ref{fig:corr_elas_const}.

According to Eq.~(\ref{eq:long_sep_form}), the growth of the
correlation function $\gfunc$ for large $\left| \xv \right|$ is
logarithmic with prefactor proportional to $g_0 \xi^\phi$, where
$\phi = 2$ for the Lorentzian case we are now considering. The
quantity $g_0 \xi^{\phi}$ is equivalent to $\Delta^K(\qv\!=\!0) $,
but this quantity is difficult to measure numerically. However,
$g_0$ can also be expressed in terms of the coordinate space
correlation $\Delta^K(\xv\!=\!0)$. The latter quantity is easily
measured by averaging $\left< \left( \delta K_{xyxy}(\xv)
\right)^2 \right>$ over all nodes. For the form of the correlation
function $g(u)$ given in Eq.~(\ref{eq:g(u)_1}),
\begin{equation}
\Delta^K(\xv\!=\!0)  \sim  g_0 \xi^{\phi-2}
\begin{cases}
\frac{1}{\phi-2}
\left(1-\left(1+\left(\Lambda \xi\right)^2\right)^{1-\frac{\phi}{2}}\right)
 & \phi \ne 2, \\
\frac{1}{2} \ln \left(1+\left(\Lambda \xi\right)^2\right) & \phi =
2 .
\end{cases}
\label{eq:dkq_to_dkx}
\end{equation}
 In the limit $\xi
\Lambda \gg 1$, $g_0 \sim \Delta^K(\xv\!=\!0)$ and the large
separation form of the correlation function $\gfunc$ is
logarithmic with prefactor $\gamma^2
\Delta^K(\xv\!=\!0)\xi^\phi/\mu^2$.

\begin{figure}

\includegraphics[width=2.5in]{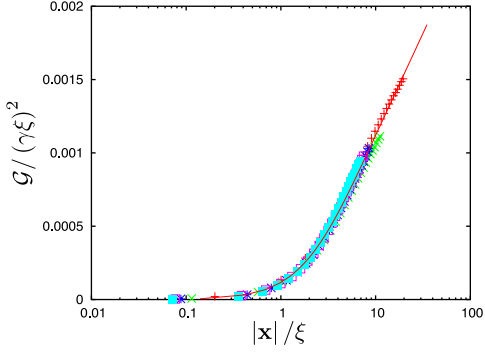}

\caption{Plot of $\gfunc/\xi^2$ versus $\xsep/\xi$ for different
$\xi$ from (points) numerical minimizations and (solid line) the
analytical expression of Eq.\ (\ref{eq:calF_I}) with a Lorentzian
$g(y)$. [Color online]}
\label{fig:corr_plot}
\end{figure}

\begin{figure}

\includegraphics[width=2.5in]{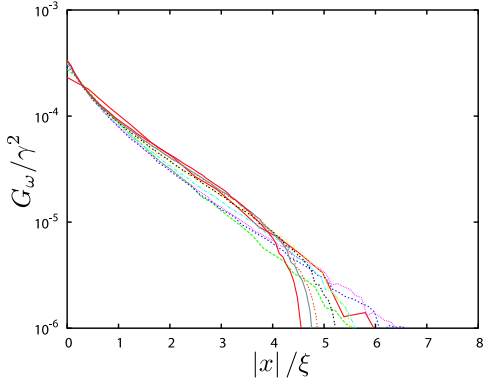}

\caption{Plot of the vorticity correlation
$G_\omega$ versus $\xsep/\xi$ for several different $\xi$ for
lattices with Lorentzian spatial correlations in $\delta k_b$.
[Color online] }
\label{fig:vort_plot}
\end{figure}

We have already established that for this set of simulations,
$\Delta^K(\xv\!=\!0)/\mu^2$ is a constant, independent of $\xi$
[see Fig.\ref{fig:corr_elas_const}]. In Fig.\
\ref{fig:corr_plot}, we plot $\gfunc/\xi^2$ versus $\xsep/\xi$ for
different values of $\xi$.  We also plot the function
$\mathcal{F}(\xsep)$ calculated from Eq.\ (\ref{eq:calF_I}) with a
Lorentzian $g(y)$ [Eq.\ (\ref{eq:Lorentzian})]. The agreement
between the numerical and analytical results is excellent with
both showing $\xsep^2$ behavior for $\xsep<\xi$ and $\ln \xsep$
behavior for $\xsep> \xi$. In Fig.~\ref{fig:vort_plot} we plot the
vorticity correlation function $G_\omega$ versus separation
rescaled by the correlation length, $\xsep/\xi$. The vorticity
correlation function decreases exponentially away from zero
separation with a decay length $\sim 1.1 \times \xi$; our
framework predicted decay with an exponent of $\xi$ exactly. The
slight discrepency between theory and simulation is not
understood.

\subsection{Model B: Internal stresses}
\label{sec:strsim}

In this model, random stresses are introduced in a periodic
lattice via a random distribution of rest bond lengths. We study
hexagonal lattices in which the rest lengths of the bonds are
multiplied by a factor $(1 + \beta_b)$ where $\beta_b$ is chosen
randomly from a flat distribution lying between $-\beta$ and
$\beta$ with $\beta<0.1$. Once again, the spring constants are set
to $k_b = 1+\delta k_b$, with $\delta k_b$ chosen randomly
from a flat
distribution lying between $-\delta \kbar$ and $+\delta \kbar$.
After specifying the rest length of each bond, we numerically
determined the equilibrium state of this random lattice with zero
applied stress by minimizing the rest energy over lattice
positions and the size of the simulation box (for a system of
40,000 particles, the minimization over box size was only a
fraction of a percent). The resulting equilibrium configuration
has zero net force on each node. This relaxed state constitutes
the reference state of our random system with lattice positions
$\Rv_{\ellv 0} = \xv $.

\begin{figure}

\includegraphics[width=2.5in]{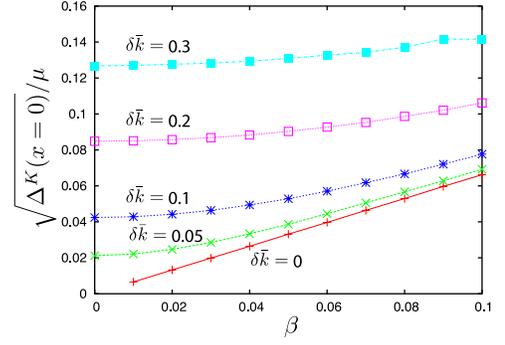}

\caption{The variance of the local shear modulus versus $\beta$ for
several different $\delta\bar{k}$. [Color online]}
\label{fig:random_stress_el}
\end{figure}

The original lattice before relaxation is characterized by random
stresses ${\overline \sigma}_{ij}$, which can be calculated from
Eq. (\ref{eq:stress_tensor_1}),
\begin{equation}
{\overline \sigma}_{ij}(\ellv) = \frac{1}{2 v} \sum_{\ellv'}
R_{bIi} R_{bIj} k_b \delta R_b /a,
\end{equation}
where $\Rv_{bI}$ is the bond vector of length $a$ (independent of
$b$) for bond $b$ in the initial undistorted hexagonal lattice and
$\delta R_b = a - R_{bR} = \beta_b a$.  The average over of
${\overline \sigma}_{ij}$ over $\beta_b$ is zero: $\langle
{\overline \sigma}_{ij} \rangle_b = 0$, and its variance is
\begin{equation}
\langle {\overline \sigma}_{ij}(\ellv) {\overline
\sigma}_{kl}(\ellv) \rangle_b = \frac{\beta^2}{27}
(1+\frac{(\delta \overline{k})^2}{3}) (\delta_{ij} \delta_{kl} +
\delta_{ik} \delta_{jl} + \delta_{il} \delta_{jk} ) .
\end{equation}
As discussed in Sec. \ref{sec:ran_stress_1}, randomness in
${\overline \sigma}_{ij}$ generates a random elastic moduli in the
relaxed reference lattice. Figure \ref{fig:random_stress_el} shows
how the random stress broadens the distribution of local elastic
moduli. For lattices with $\delta \bar{k} =0$,
$\sqrt{\Delta^K(\xv\!=\!0)}$ is linearly proportional to $\beta$
as predicted by Eqs.\ (\ref{eq:u_sigma}) to (\ref{eq:vij}).

\begin{figure}

\includegraphics[width=2.5in]{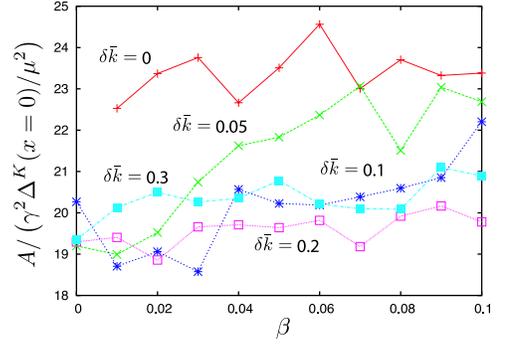}

\caption{$A/(\gamma^2 \Delta^K(\xv\!=\!0)/\mu^2)$ versus $\beta$
for several different $\delta\bar{k}$. [Color online]}
\label{fig:random_stress_G}
\end{figure}

After constructing the relaxed state, we sheared it in the $xy$
plane as before and measured the nonaffinity correlation function.
The measurements were well fit by the functional form $\gfunc \sim
A \ln (\xsep/B)$. Figure~\ref{fig:random_stress_G} shows that for
$\delta \bar{k}=0$ the measured ratio
$A/(\gamma^2 \Delta^K(\xv\!=\!0)/\mu^2)$ is nearly
independent of $\beta$, as predicted in
Section~\ref{sec:ran_stress_1}. For $\delta \bar{k}>0$, the ratio
$A/(\gamma^2 \Delta^K(\xv\!=\!0)/\mu^2)$ is $\sim 20\%$ lower at small $\beta$, but
asymptotes to the $\delta \bar{k}=0$ value as $\beta$ increases,
approaching the asymptote more quickly for smaller $\delta
\bar{k}$. The difference in
$A/(\gamma^2 \Delta^K(\xv\!=\!0)/\mu^2)$ between stressed and
stress-free lattices is most likely a higher order effect due to
the breaking of hexagonal symmetry as $\beta$ is
increased.

\subsection{Model C: Random lattice}
\label{sec:ransim}

In this model, the initial reference lattice is geometrically
random. The rest bond length $R_{bR}$ is equal to the equilibrium
lattice parameter $R_{b0}$ for every bond, so that the reference
state is stress free. Our method of generating reference lattices
of varying randomness is detailed below. Each bond is assigned the
anharmonic potential of Eq.~(\ref{eq:anharpot}), where the spring
constant $k_b(R_b) = k_0/R_b$ is a constant {\em per unit length}
of the rest bond length.

We use the approach followed in~\cite{ChungGie2002} to generate
networks with a tunable degree of randomness. We begin by
simulating a $2$-dimensional gas of $40000$ point particles
interacting through a Lennard-Jones potential. The procedure
outlined in~\cite{BerendsenHaa1984} is used to equilibrate the gas
at a prescribed temperature and pressure, with periodic boundary
conditions. The gas is equilibrated for $10000$ time steps, after
which the particle configurations are sampled every $1000$  time
steps. In this manner we obtain $40$ uncorrelated configurations
of the gas at thirteen different temperature-pressure combinations,
with $T=8.0$ and $P=0.025$, $0.05$, $0.1$, $0.2$, $0.25$, $0.3$,
$0.35$, $0.4$, $0.5$ $0.6$, $0.7$, $0.8$, and
 $1.0$, all in units of the Lennard-Jones potential.

\begin{figure}

\includegraphics[width=2.5in]{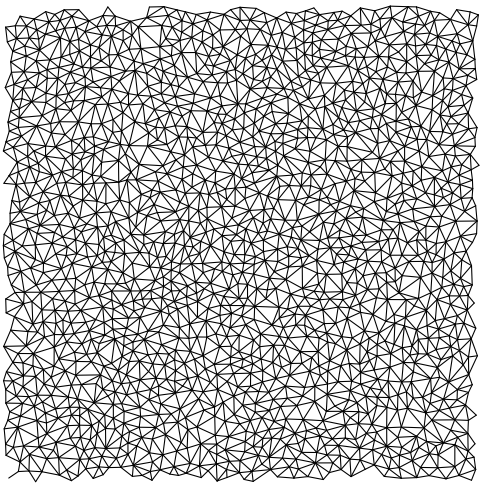}

\caption{Section of a random lattice created by triangulating a
snapshot of a Lennard-Jones gas with $T=8.0$ and $P=0.05$.}
\label{fig:ranlat}
\end{figure}







We use the particle positions from the snapshots of the
equilibrated gas as the positions of nodes in our random lattice.
Each sampled configuration is rescaled to have a box length of $1$
on each side. The point configurations are then tesselated using
the Delaunay triangulation, which places a bond between each node
and its nearest neighbors. The Delaunay triangulation produces
networks with an average of $6$ bonds per node. A resulting
lattice is pictured in Fig.~\ref{fig:ranlat}.

The randomness in local elastic moduli as calculated from
Eq.~(\ref{eq:local_el_modulus}) is proportional to the
distribution of bond lengths and bonds per node.
In principle, as we take the equilibrium
gas pressure to zero, the distribution of bond lengths will become
completely random. Conversely, as we increase the pressure past a
critical point the simulated gas begins to crystalize, forming
spatial domains of hexagonal order separated by grain boundaries.
This transition should be marked by a growth in the two-point
correlation of local shear moduli. We fit the non-affinity
correlation data for a broad range of pressures which cross this
transition and compare it to the framework developed in previous
sections.
We used Eq.~(\ref{eq:local_el_modulus})
to calculate $\Delta^K(\xv\!=\!0)/\mu^2$ for each ensemble
of random lattices, while the crystalline correlation length
is fit as an unknown.

\begin{figure}

\includegraphics[width=2.5in]{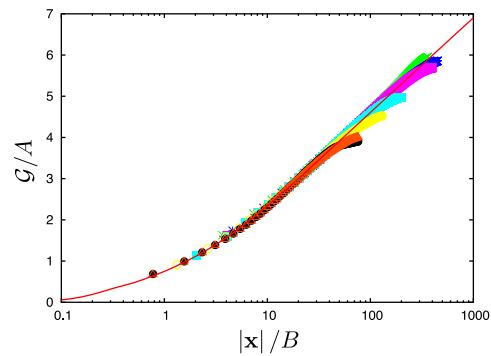}

\caption{$\gfunc$ vs $\xsep$ for lattices
with varying degrees of geometrical randomness, scaled to have the
same asymptotic form for large $\xsep$. The solid line is a fit
Eq.~(\ref{eq:calF_I}) using $g(u)$ of Eq.~(\ref{eq:g(u)_1}) with
$\phi = 0.4$. [Color online] }
\label{fig:radg}
\end{figure}

The lattice is sheared by $0.1 \%$ and the energy is minimized as
a function of node position as before. Figure~\ref{fig:radg} shows
the displacement correlation function $\gfunc$ as a function of
separation $\xsep$ for lattices with different degrees of
randomness. This correlation function shows the same logarithmic
growth at large $\xsep$ as it does in the random spring constant
lattices from the last section.

\begin{figure}

\includegraphics[width=2.5in]{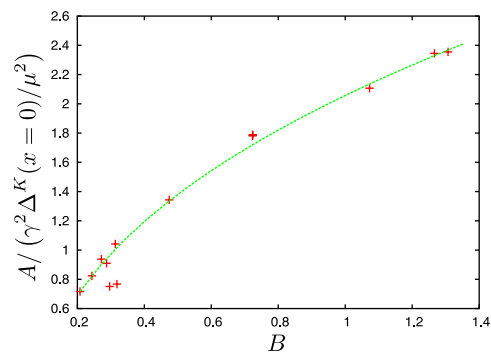}

\caption{$A/(\gamma^2 \Delta^K(\xv\!=\!0)/\mu^2)$ versus $B$ for
geometrically random lattices, where $A$ and $B$ are the fit
parameters to the functional form $\gfunc= A \ln \left( \xsep / B
\right)$. The line is a fit to these data points using
Eq.~(\ref{eq:abfit}), assuming that there are long range
correlations in the elastic moduli. [Color online]}
\label{fig:random_abfit}
\end{figure}

We fit the measurements of $\gfunc$ to the functional form $A \ln
\left( \xsep / B \right)$ at large $\xsep$. This data is shown in
Figure~\ref{fig:random_abfit}. For the very random lattices
generated at low Lennard-Jones pressure ($T=8.0$, $P < 0.3$) the
values of $A/(\gamma^2 \Delta^K(\xv\!=\!0)/\mu^2) $ and $B$ are
nearly constant, as our framework predicts for the simple case of
delta-function spatial correlations. However, lattices created at
higher pressure values ($T=8.0$, $P \ge 0.3$) showed significant
growth of both $A/(\gamma^2 \Delta^K(\xv\!=\!0)/\mu^2) $ and $B$
with increasing pressure, reaching a saturation point at around
$P=8.0$. Visual inspection of the lattices in question revealed
subdomains of hexagonal crystalline ordering. Long range
correlations in the connectivity implies long-range correlations
in the elastic moduli, so we must apply the framework developed in
Sec.\ \ref{sec:long-range-corr} and App.\ \ref{App:C1} in order to
fit the data for partially crystalline lattices. Once again, we
try the functional form in Eq.~(\ref{eq:g(u)_1}) for the spatial
correlations in the elastic modulus. The numerical value of the
factor $g_0 (\phi)$ can be calculated from the measured modulus
autocorrelation $\Delta^K(\xv\!=\!0)$ using
Eq.~(\ref{eq:dkq_to_dkx}).

The fitting line in Fig.~\ref{fig:random_abfit} represents a best
fit of both the correlation exponent $\phi$ and the cutoff length
$\Lambda^{-1}$ to the form
\begin{align}
A &\sim \frac{\Delta^K(\xv\!=\!0) \gamma^2}{\mu^2}
\ \xi^{2}
\left(1-\left(1+\left(\Lambda \xi\right)^2\right)^{1-\frac{\phi}{2}}
\right)^{-1} \nonumber \\
B &\sim \frac{\xi}{\beta(\xi\Lambda,\phi)},
\label{eq:abfit}
\end{align}
as suggested by Eq.\ (\ref{eq:long_sep_form}). The best fit was
achieved for $\phi=0.4$ and a cutoff length of
$\Lambda^{-1}\approx 1.25$ lattice spacings. The corresponding
analytic form of $\gfunc$ calculated from Eq.~(\ref{eq:calF_I})
using $g(u)$ from Eq.\ (\ref{eq:g(u)_1}) is shown by the solid
line in Fig.~\ref{fig:radg}.

\begin{figure}
\centerline{\includegraphics[width=2.5in]{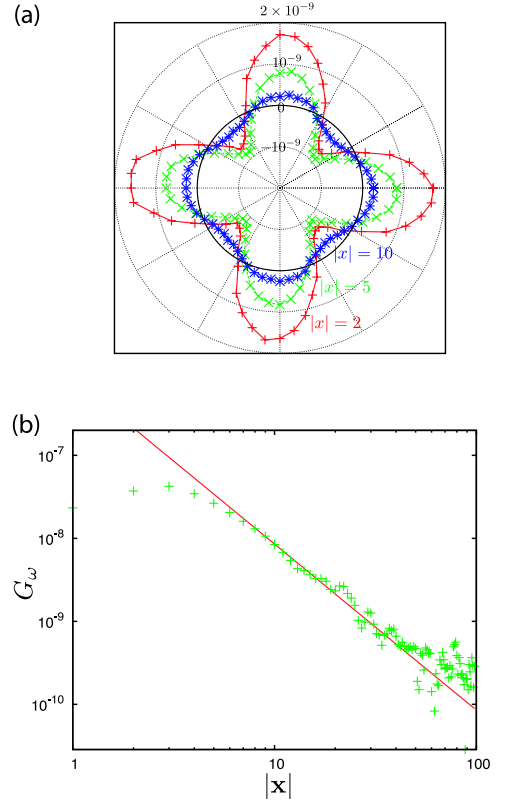}}

\caption{(a) Polar plot showing the $\cos 4 \psi$ modulation of the
vorticity correlation function $G_{\omega} ( \xv)$ for $\xsep =
2$, $5$, and $10$ lattice spacings. (b) A log-log plot of the
coefficient of $\cos 4 \psi$ in $G_{\omega} ( \xv)$ showing the
expected $|\xv|^{-2}$ fall off at large $|\xv|$. [Color online]}
\label{fig:vortcor4}
\end{figure}

To test the predicted [Eq.\ (\ref{eq:vortex_corr})] $\cos 4 \psi$
anisotropy in vorticity correlations, we measured $G_{\omega} (
\xv )$ as a function of the angle $\xv$ makes with the $x$-axis.
Figure \ref{fig:vortcor4} shows a polar plot of $G_{\omega} ( \xv
)$, which clearly shows $\cos 4 \psi$ behavior, and the dependence
of the $\cos 4 \psi$ term on $|\xv|$, which shows the expected
$|\xv|^{-2}$ behavior.

\subsection{Model D: Random lattice with internal stresses}
\label{sec:ranstrsim}

\begin{figure}

\centerline{\includegraphics[width=2.5in]{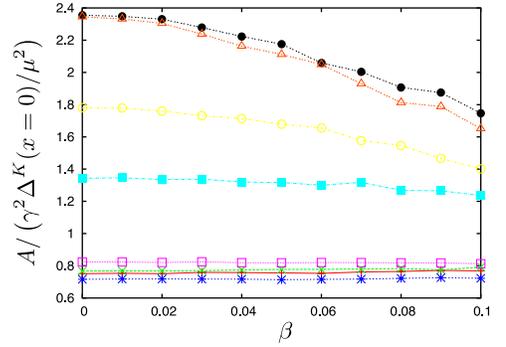}}

\caption{
$A/(\gamma^2 \Delta^K(\xv\!=\!0)/\mu^2)$ versus $\beta$ for
geometrically random lattices with internal stresses, where $A$ is
the coefficient of a logrithmic fit to the non-affinity
correlation function $\gfunc$ for each lattice and $\beta$ is the
bond frustration factor used to induce internal stresses. The
lines are for lattices with eight different temperature-pressure
combinations, with $T=8.0$ and, from bottom to top $P=0.025$,
$0.05$, $0.1$, $0.2$, $0.4$, $0.6$, $0.8$, and $1.0$, all in units
of the Lennard-Jones potential. [Color online]}
\label{fig:random_str}
\end{figure}

Finally, we simulate the most general model for random lattices,
in which the rest bond length $R_{bR}$ is not equal to the
equilibrium lattice parameter $R_{b0}$, and the lattice parameters
$R_{b0}$ along with the number of bonds per node are random to
within some finite distribution. Each bond is assigned the
anharmonic potential of Eq.\ (\ref{eq:anharpot}), where the spring
constant $k_b(R_b) = k_0/R_b$ is a constant {\em per unit length}
of the rest bond length. We used the same geometrically random
lattices from Section~\ref{sec:ransim} as staring points, then we
add bond length frustration using the technique from
Section~\ref{sec:strsim}: We multiply the rest lengths of all
bonds by a factor $(1 + \beta_b)$ where $\beta_b$ is chosen
randomly from a flat distribution lying between $-\beta$ and
$\beta$ with $\beta<0.1$. We find the equilibrium configuration of
the lattice by minimizing the elastic energy over node positions
and box size. We then shear the lattice by $0.1\%$, minimize the
energy over node positions, and measure the non-affinity
correlation function $\gfunc$.

In all these simulations, the correlation function $\gfunc$ was
well fit by the functional form $A \ln \left(\xsep / B \right)$.
Figure~\ref{fig:random_str} shows a plot of $A/(\gamma^2 \Delta^K(\xv\!=\!0)/\mu^2)$
for all data sets as a function of $\beta$. The data points for $\beta=0$ correspond to the
data from Section~\ref{sec:ransim}; their deviation from the
expected constancy of $A/(\gamma^2 \Delta^K(\xv\!=\!0)/\mu^2)$ was explained in
that section by the growth of a correlation length scale as the
system acquires partial hexagonal crystalline ordering. Here we
see that as $\beta$ is increased, the long length scale ordering
is disrupted by the additional randomness, and the ratio
$A/(\gamma^2 \Delta^K(\xv\!=\!0)/\mu^2)$ decreases toward the value for
completely disordered lattices.

\section{Summary and Conclusions}

Nonaffine distortions are always present in random elastic
networks subjected to external stress.  In this paper, using both
analytical and numerical techniques, we study properties of
nonaffinity in these systems manifested in correlation functions
of the deviation, $\uv'(\xv)$, of local displacements from their
affine form. We introduce four models of random elastic networks
with random local elastic moduli and possibly local random stress
arising either from randomness in the form of the central force
potentials between nearest neighbor sites or from random
connectivity of the the network. In all cases, we show
analytically and verify with numerical simulations that random
elastic modulus times imposed strain and not random stress act as
sources for nonaffine distortions.  We calculate the nonaffinity
displacement correlation function, $\gfunc = \langle [ \uv'( \xv)
- \uv'(0)]^2\rangle$, and the vorticity correlation function,
$G_\omega(\xv) = \langle \omega ( \xv ) \omega ( 0 )\rangle$
analytically and verify their form in numerical simulations for
systems with both short- and long-range correlations in local
elastic moduli.  We show in particular that $\gfunc \sim \gamma^2
(\langle(\delta K)^2 \rangle/K^2) \ln |\xv|$ at large $\xv$ in two
dimensions, where $\gamma$ is the imposed strain, $K$ is the
average of elastic modulus, and $\langle(\delta K)^2 \rangle$ is
it variance.

The formalism we develop is general and should be applicable to
any elastic system that has a well defined average shear modulus.
It should provide a basis for studying nonaffinity in granular
media, foams, networks of semi-flexible polymers, and related
systems. It should, in particular, provide a method of calculating
correlation lengths near percolation-like thresholds such as the
$J$-point in jammed systems or the rigidity percolation point. We
have begun \cite{VernonLub2005}  to use these techniques to
calculate correlation lengths in the former systems which we will
eventually compared with those calculated from the density of
states \cite{WyartWit2005,SilbertNag2005} and to study nonaffinity
in networks of semi-flexible polymers \cite{DidonnaLev2005}.

\begin{acknowledgments}
We are grateful to Peter Sollich and Dan Vernon for careful
readings of the manuscript and their resultant useful suggestions
and identification of misprints. BD gratefully acknowledges
helpful discussions with Eric van der Giessen, Mitchell Luskin,
Fred Mackintosh and Michael Rubinstein. This work was supported in
part by the National Science Foundation under DMR 04-04670 (TCL),
the National Institutes of Health under grant R01 GM056707 (BD and
TCL), and the Institute for Mathematics and its Applications with
funds provided by the National Science Foundation.
\end{acknowledgments}

\appendix
\section{\label{App:A}Properties of the modulus correlator}
The  modulus correlator $\Delta^K_{ijkl;i'j'k'l'}(\qv)$ is an 8th
rank tensor.  The number of its independent components depends on
the symmetry of the reference space.  In this appendix, we will
determine the number and form of its independent components at
$\qv=0$ (strictly speaking $\qv \rightarrow 0$), or, equivalently,
at all $\qv$ when correlations are short range and it is
independent of $\qv$, when the reference space is isotropic. In
this case, the general form of $\Delta^K_{ijkl;i'j'kj'l'} \equiv
\Delta^K_{ijkl;i'j'k'l'}(\qv=0)$ must be constructed from products
of Kroneker $\delta$'s that distinctly pair all indices while
respecting all symmetries.

It is useful to recall how this process is carried out for the
simpler case of the 4th-rank elastic-modulus tensor $K_{ijkl}$,
which is symmetric under interchange of $i$ and $j$, of $k$ and
$l$, and of the pairs $ij$ and $kl$.  Since any index can be
paired with any of the remaining three and there is only one way
to pair the remaining two, there are three distinct
Kroneker-$\delta$ pairings, which we will call contractions, of
the four indices: $\delta_{ij}\delta_{kl}$,
$\delta_{ik}\delta_{jl}$, and $\delta_{il}\delta_{jk}$.  The first
of these satisfies all of the symmetry constraints, but the second
two do not;  their sum, however, does. The elastic-modulus tensor,
therefore, has two independent components in an isotropic medium:
$K_{ijkl} = \lambda \delta_{ij} \delta_{kl} + \mu (\delta_{ik}
\delta_{jl} + \delta_{il} \delta_{jk})$.

\begin{figure}
\centerline{\includegraphics{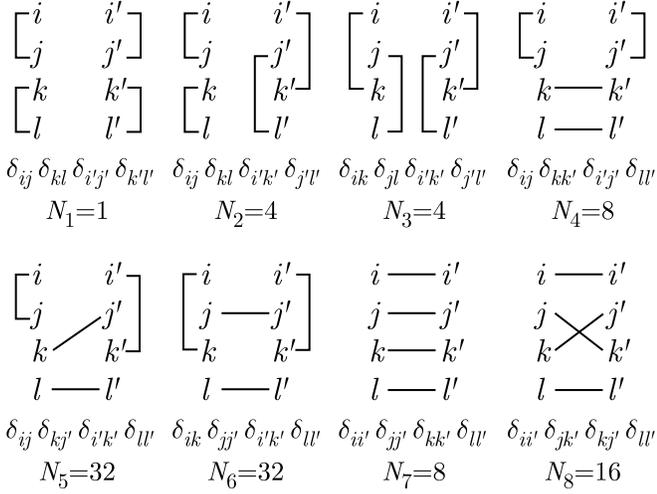}}
\caption{This figure provides a graphical representation of the
groups of contractions that are invariant under the symmetry
operations that leave $\Delta^K_{ijkl;i'j'k'l'}$ unchanged.  It
shows 8 graphs representing contractions, none of which are
transformed into any of the others under any symmetry operation.
The symmetry operations applied to each contraction will, however,
generate other contractions.  The number of contractions for each
graph produced by performing all symmetry operations on it is
indicated below each graph along with a representation of the
graph in terms of Kronecker $\delta$'s.}
\label{fig:Delta-contr}
\end{figure}

$\Delta^K_{ijkl;i'j'k'l'}$ is symmetric under interchange of $i$
and $j$, $k$ and $l$, $i'$ and $j'$, and $k'$ and $l'$; under the
interchange of the pairs $ij$ and $kl$ and of the pairs $i'j'$ and
$k'l'$; and under the interchange of the four-plets $ijkl$ and
$i'j'k'l'$.  The total number of possible contractions of these 8
indices is $N_T = 7 \times 5 \times 3 \times 1 = 105$ because any
index can be contracted with any of the seven remaining indices,
any one of the six remaining indices can then be contracted with
any of the other five remaining, etc. Most of the individual
realizations of these 105 possible contractions will not satisfy
symmetry constraints; it is necessary to find the linear
combinations of them that do.  Figure \ref{fig:Delta-contr}
provides a graphical representation of the eight distinct
contraction groups the sum over whose elements satisfy all
constraints.  The elastic-modulus correlation function in an
isotropic medium can thus be written as
\begin{eqnarray}
&&\Delta^K_{ijkl;i'j'k'l'} = \sum_\alpha \Delta_{ijkl;i'j'k'l'}^{K
\alpha} \nonumber \\
& & = \Delta_1 \delta_{ij}\delta_{kl}
\delta_{i'j'} \delta_{k'l'} \nonumber \\
&& + \Delta_2 \delta_{ij} \delta_{kl} (\delta_{i'k'} \delta_{j'l'}
+ \delta_{i'l'} \delta_{j'k'}) + \text{prime} \leftrightarrow
\text{unprime} \nonumber\\
&&+ \Delta_3 (\delta_{ik} \delta_{jl} + \delta_{il}
\delta_{jk})(\delta_{i'k'} \delta_{j'l'} + \delta_{i'l'}
\delta_{j'l'}) \nonumber \\
&&+ \Delta_4 \delta_{ij} \delta_{i'j'} \delta_{kk'} \delta_{ll'} +
\text{7 perm.} \nonumber \\
&&+ \Delta_5 \delta_{ij}\delta_{i'k'}\delta_{kj'}\delta_{ll'} +
\text{31 perm.} \nonumber \\
& & + \Delta_6 \delta_{ik} \delta_{i'k'} \delta_{jj'} \delta_{ll'}
+ \text{31 perm.} \nonumber \\
&& + \Delta_7 \delta_{ii'} \delta_{jj'} \delta_{kk'} \delta_{ll'}
+ \text{7 perm.} \nonumber \\
&& + \Delta_8 \delta_{ii'} \delta_{jk'} \delta_{kj'} \delta_{ll'}
+ \text{15 perm.}
\label{eq:DeltaKa}
\end{eqnarray}
where $\Delta^{K1}_{ijkl;i'j'k'l'} = \Delta_1
\delta_{ij}\delta_{kl} \delta_{i'j'} \delta_{k'l'}$, etc. The
first three terms in $\Delta^{K}_{ijkl;i'j'k'l'}$ describe
correlations in the isotropic Lam\'{e} coefficients: $\Delta_1 =
\langle (\delta \lambda)^2 \rangle$, $\Delta_2 = \langle \delta
\lambda \delta \mu \rangle$, and $\Delta_3 = \langle (\delta
\mu)^2 \rangle$.  The other terms represent fluctuations away from
local isotropy.

\section{\label{App:B}Evaluation of $G(\qv)$}

We outline here the calculation of $G(\qv)$ to lowest order in
$\Delta^K$ in isotropic systems.  We use Eq.\ (\ref{eq:full_G})
for $G_{ij}(\qv)$ and sum over $i=j$.  Using Eq.\ (\ref{eq:chi0}),
we find
\begin{equation}\
\chi_{ip}^0 ( \qv ) \chi_{ip'}^0 ( \qv ) = \frac{1}{\mu^2
q^4}\delta_{pp'}^T + \frac{1}{(\lambda + 2 \mu)^2 q^4} \hatq_p
\hatq_{p'} ,
\end{equation}
where $\hatq_p = q_p/q$ and $\delta_{pp'}^T = \delta_{pp'} - \hatq_p
\hatq_{p'}$. Then
\begin{equation}
G(\qv) = \gamma_{xy}^2 \sum_{\alpha=1}^8 \left(\frac{1}{\mu^2
q^2}S_{\alpha}^T + \frac{1}{(\lambda + 2 \mu)^2
q^2}S_{\alpha}^L\right) ,
\end{equation}
where
\begin{equation}
S_{\alpha}^T = \delta_{pp'}^T S_{\alpha pp'}\qquad S_{\alpha}^L =
\hatq_p \hatq_{p'} S_{\alpha pp'} ,
\end{equation}
with
\begin{equation}
S_{\alpha p p'} = \Delta^{K \alpha}_{pjxy;p'j'xy} \hatq_j
\hatq_{j'}
\label{eq:Salphapp'}
\end{equation}
where $\Delta^{K\alpha}_{ijkl;i'j'k'l'}$ is defined in Eq.\
(\ref{eq:DeltaKa}).  It is straightforward but tedious to
calculate $S_{\alpha}^T$ and $S_{\alpha}^L$ from Eq.\
(\ref{eq:DeltaKa}).  The results are
\begin{equation}
\begin{array}{ll}
S_1^T = S_2^T = 0, & S_1^L = S_2^L = 0; \\
S_3^T = \Delta_3^T(\hatq_{\perp}^2 - 4 \hatq_x^2 \hatq_y^2), &
S_3^L
=\Delta_3^L \hatq_x^2 \hatq_y^2; \\
S_5^T = 0, & S_5^L = \Delta_5^L 4 \hatq_{\perp}^2  \\
S_6^T = \Delta_6^T(2 + d \hatq_{\perp}^2 - 16 \hatq_x^2), &  S_6^L
= \Delta_6^L (4 \hatq_{\perp}^2 + 16 \hatq_x^2 \hatq_y^2);  \\
S_7^T = \Delta_7^T[(d-1)+ \hatq_{\perp}^2-4 \hatq_x^2 \hatq_y^2];
&
S_7^L = \Delta_7^L(2 + 4 \hatq_x^2 \hatq_y^2)  \\
S_8^T = \Delta_8^T [2 + (d-2) \hatq_{\perp}^2], &S_8^L =
\Delta_8^L 4 \hatq_{\perp}^2 .
\end{array}
\end{equation}

\section{\label{App:C}Evaluation of ${\mathcal{G}(\xv)}$}

In this appendix, we will evaluate the integrals
[Eq.~(\ref{eq:calF})]
\begin{equation}
\mathcal{F}_\alpha = 2 \xi^{\phi} \int \frac{d^d q}{(2 \pi)^d}
\frac{1}{q^2} f_{\alpha}( \qv) g_{\alpha}( q \xi)\left( 1 - e^{i
\qv\cdot\xv}\right)
\end{equation}
in $2D$ and $3D$ that make up the function $\gfunc$, where $f_A
(\qv)= 1$, $f_B(\qv) = \hat{q}_{\perp}^2$, and $f_C(\qv) =
\hat{q}_x^2 \hat{q}_y^2$.

\subsection{\label{App:C1}Two dimensions}
In two dimensions, $f_B ( \qv) = 1 = f_A(\qv)$, and
\begin{equation}
f_C( \qv) = \sin^2 \phi_q \cos^2 \phi_q = \frac{1}{8}(1 - \cos 4
\phi_q),
\end{equation}
where $\phi_q$ is the angle between $\qv$ and the $x$-axis. Using
the plane-wave decomposition relation
\begin{equation}
e^{i \qv \cdot \xv} = J_0 ( q |\xv|) + 2 \sum_{n=1}^\infty \cos n
\Theta J_n ( q |\xv|),
\end{equation}
where $J_n(x)$ is the $n$th order Bessel function, $\Theta =
\phi_q - \psi$, and $\psi$ is the angle between $\xv$ and the
$x$-axis, and the orthogonality relation
\begin{equation}
\frac{1}{2 \pi}\int_0^{2 \pi}d \phi \cos n \psi \cos m \Theta =
 \begin{cases}
 \frac{1}{2} \delta_{nm} \cos n \psi & n\neq 0\\
 \delta_{nm} & n=0
 \end{cases}
,
\end{equation}
we find
\begin{equation}
\mathcal{F}_\alpha  \equiv F_I [g_\alpha] =
\frac{\xi^{\phi}}{\pi}\int_0^{\Lambda} \frac{dq}{q} g_\alpha( q
\xi) [ 1 - J_0 ( q |\xv|)]
\label{eq:FI}
\end{equation}
for $\alpha = A,B$ and
\begin{equation}
\mathcal{F}_C = \mathcal{F}_I[g_C] - \cos 4 \psi F_A[g_C] ,
\end{equation}
where
\begin{equation}
F_A[g_\alpha] = \frac{\xi}{\pi} \int_0^{\Lambda} \frac{dq}{q}
g_\alpha ( q \xi) J_4 ( q |\xv|) .
\end{equation}
Thus,
\begin{equation}
\mathcal{F}( \xv ) = F_I [g] + \frac{1}{8} \cos 4 \psi F_A[g_C]
,
\end{equation}
where
\begin{equation}
g(q \xi ) = g_A ( q \xi) + g_B ( q \xi) - \frac{1}{8} g_C( q \xi )
.
\end{equation}

We now evaluate the integrals $F_I$ and $F_A$ in the limits
$|\xv|\gg \xi \geq \Lambda^{-1}$ and $\Lambda^{-1} \ll |\xv| \ll
\xi$.

\subsubsection{$|\xv| \gg \xi \geq \Lambda^{-1}$ in Two Dimensions}

To evaluate the first limit of $F_I$, we set $y = q |\xv|$ in Eq.\
(\ref{eq:FI}):
\begin{align}
& F_I [g] = \frac{\xi^{\phi}}{\pi} \Big\{ \int_1^{\Lambda |\xv|}
\frac{dy}{y} g( y \xi/|\xv|) \nonumber\\
& +\int_0^1 \frac{dy}{y} g( y \xi/|\xv|)[1- J_0 ( y)]  -
\int_1^{\Lambda |\xv|} \frac{dy}{y} g( y \xi/|\xv|) J_0(y) \Big\}.
\label{eq:F_I_2d}
\end{align}
In the limit $|\xv|/\xi \rightarrow \infty$, we can safely replace
$g( y \xi/|\xv|)$ by $g_0$ in the second and third integrals in
this expression, and we can let $\Lambda |\xv| \rightarrow \infty$
in the third integral.  The first integral diverges as $\ln |\xv|$
if we replace $g( y \xi/|\xv|)$ by $g_0$ in it, and we have to be
more careful to extract the constant term beyond the log:
\begin{align}
\int_1^{\Lambda |\xv|}&\frac{dy}{y} g( y \xi/|\xv|)  =
\int_{\xi/|\xv|}^1 \frac{du}{u} g(u) + \int_1^{\Lambda \xi}
\frac{du}{u} g(u) \\
& \rightarrow g_0 \ln |\xv|/\xi + \int_0^1\frac{du}{u} [g(u) -
g_0] + \int_1^{\Lambda \xi} \frac{du}{u} g(u).
\end{align}
Thus in the limit $|\xv| \gg \xi \geq \Lambda^{-1}$,
\begin{equation}
F_I [g] = \frac{\xi^{\phi}}{\pi} g_0 \ln \beta(\Lambda \xi,\phi)
\frac{|\xv|}{\xi} ,
\label{eq:long_sep_form}
\end{equation}
where
\begin{equation}
\ln \beta(\Lambda\xi,\phi) = \ln \alpha + \int_0^1 \frac{dy}{y}
\left[\frac{g(y)}{g_0}-1\right]+ \int_1^{\Lambda \xi }
\frac{dy}{y} \frac{g(y)}{g_0} ,
\end{equation}
where
\begin{equation}
\ln \alpha = \int_0^1\frac{dy}{y}[1-J_0 ( y)] - \int_1^{\infty}
\frac{dy}{y} J_0 (y)  = -0.1159
\end{equation}
and $\alpha = 0.8905$. When $g(y) = g_0$, independent of $y$,
\begin{equation}
\ln \beta( \Lambda \xi ) \rightarrow \ln \alpha + \ln \Lambda \xi
,
\end{equation}
and Eq.\ (\ref{eq:F_I_2d}) reduces to Eq.\ (\ref{eq:calG_2d}) when
$g_0 \xi^\phi$ is identified with $\Delta^S$.

The $|\xv| \gg \xi$ limit of $F_A$ is obtained by setting $y = q
|\xv|$ and noting that letting the upper limit of the integral go
to infinity and replacing $g(y |\xv|/\xi)$ by $g_0$ introduces no
singularities.  The result is
\begin{equation}
F_A [g] = \frac{\xi}{\pi} g_0 \int_0^{\infty} \frac{dy}{y} J_4 ( y
)= \frac{g_0 \xi}{4 }.
\end{equation}

\subsubsection{$\Lambda^{-1} \ll |\xv| \ll \xi$ in Two Dimensions}

To evaluate integrals when $\Lambda^{-1} \ll |\xv| \ll \xi$, we
introduce a new function
\begin{equation}
h(u) = u^{\phi} \frac{g(u)}{g_\infty} \sim
 \begin{cases}
 1 & \text{as $u\rightarrow \infty$},\\
 (g_0/g_\infty) u^\phi & \text{as $u\rightarrow 0$} ,
 \end{cases}
\label{eq:h(u)}
\end{equation}
where $g_{\infty}$ is defined in Eq.~(\ref{eq:ginfty}) Then
$\mathcal{F}_I[g] = g_{\infty}(\xi^\phi/\pi)(|\xv|/\xi)^{\phi}
\mathcal{I} ( \xv)$, where
\begin{equation}
\mathcal{I}=\int_0^{\Lambda|\xv|} \frac{dy}{y} y^{-\phi}h(y
\xi/|\xv|)[1-J_0 ( y)] .
\end{equation}
This integral has a potential infrared divergence as $\xi/|\xv|
\rightarrow \infty$ when $\phi \geq 2$.  To isolate it, we break
up the integral from $0$ to $\Lambda |\xv|$ into one from $0$ to
$1$ and another from $1$ to $\Lambda |\xv|$.  There are no
troubles with ultraviolet divergences in the second integral, and
in it, we can let $\Lambda |\xv| \rightarrow \infty$ and replace
$h(y \xi/|\xv|)$ by its infinite argument limit of one.  In the
integral from $0$ to $1$, we extract the small $y$ behavior of
$1-J_0(y)$ via $1-J_0(y) = (y^2/4) + [1 - J_0(y) -(y^2/4)]$.  The
second part of this expression vanishes as $y^4$ at small $y$, and
there is no infrared divergence in the integral involving it so
long as $\phi <4$. Thus, we have
\begin{align}
\mathcal{I} = \mathcal{I}_1 & + \int_0^{1} \frac{dy}{y}
y^{-\phi}[1-J_0(y) - (y^2/4)]\\
& + \int_1^{\infty} \frac{dy}{y} y^{-\phi}[1-J_0(y)] ,
\end{align}
where
\begin{align}
\mathcal{I}_1 & =\frac{1}{4}\int_0^1 dy y^{1-\phi} h(y\xi/|\xv|),
\\
& =
\frac{1}{4}\left(\frac{\xi}{|\xv|}\right)^{\phi-2}\int_0^{\xi/|\xv|}
du u^{1-\phi} h(u) \\
&=\frac{1}{4}\left(\frac{\xi}{|\xv|}\right)^{\phi-2}\Big\{\int_0^1
du
u^{1-\phi} h(u) \\
& \qquad + \int_1^{\xi/|\xv|} dy u^{1-\phi} + \int_1^{\xi/|\xv|}
u^{1-\phi} [h(u) - 1] \Big\} .
\end{align}
Using $\int_1^\eta u^{1-\phi} = [\eta^{2-\phi} -1]/(2-\phi)$, we
arrive at Eq.\ (\ref{eq:2D_correlated}) with
\begin{align}
\mathcal{A}_2(\phi) & = \int_0^{\infty} dy y^{-(1+\phi)}
[1-J_0(y)] \\
\ln \nu & = \int_0^1 \frac{du}{u} h(u) + \int_1^{\infty}
\frac{du}{u} [h(u) -1] \label{eq:nu}\\
\mathcal{C}_2(\phi)& = \int_0^\infty du u^{1-\phi} h(u) .
\label{eq:calC}
\end{align}

The evaluation of the $\xi \gg |\xv|$ limit of $F_A [g]$ is
straightforward.  The result is
\begin{equation}
F_A = g_\infty \frac{|\xv|^{\phi}}{\pi} \int_0^{\infty}
\frac{dy}{y} y^{-\phi} J_4(y) .
\end{equation}

\subsection{Three Dimensions}

To evaluate the integrals $\mathcal{F}_{\alpha}$ in $3D$, we make
use of the $3D$ plane-wave decomposition:
\begin{equation}
e^{i \qv \cdot \xv} = 4 \pi \sum_{l=0}^{\infty}i^l j_l ( q |\xv|)
\sum_{m=-l}^l Y_{lm}(\Omega_q ) Y_{lm}^*(\Omega_x) ,
\end{equation}
where $\Omega_q = (\theta_q, \phi_q)$ and $\Omega_x = (\theta_x,
\phi_x)$ are, respectively, the polar angles of $\qv$ and $\xv$,
$Y_{lm} ( \Omega)$ are spherical harmonics, and $j_n(u)$ is the
$n$th order spherical Bessel function. Then, noting that
\begin{align}
{\hat q}_{\perp}^2 & = \sin^2 \theta_q  = \frac{2}{3} \left[1 -
P_2 (\cos \theta_q)\right] ,\\
{\hat q}_x^2 {\hat q}_y^2 & = \frac{1}{8} \sin^2 \theta_q ( 1 -
\cos 4 \phi_q ) \\
& = \frac{1}{105}[7 - 10 P_2 ( \cos \theta_q) + 3 P_4 ( \cos
\theta_q )] \\
& \,\,\,- \frac{1}{16} \sqrt{\frac{4 \pi}{9!}}2^4 4!
[Y_{44}(\Omega_q) + Y_{4,-4}(\Omega_q) ] ,
\end{align}
where $P_n(x)$ is the $n$th-order Legendre Polynomial, we find
\begin{align}
\mathcal{F}_A & = \frac{\xi^\phi}{\pi} I_0[g_A] \\
\mathcal{F}_B& = \frac{\xi^{\phi}}{\pi}\left\{\frac{2}{3}I_2[g_B]
+ \frac{2}{3}P_2 ( \cos \theta_x) I_2[g_B] \right\} \\
\mathcal{F}_C& =\frac{\xi^{\phi}}{\pi}\Big\{\frac{1}{15} I_0[g_C]
-\frac{2}{21}P_2(\cos \theta_x) I_2[g_C] \\
& \,\,\, - \left[ \frac{1}{35}P_2(\cos\theta_x) + \frac{1}{8}
\sin^2 \theta_x \cos 4 \phi_x \right] I_4 [g_C] ,
\end{align}
where
\begin{align}
I_0[g] & = \int_0^{\Lambda} dq g(q \xi) [ 1 - j_0 ( q |\xv|)]\\
I_2[g] & = \int_0^{\Lambda} dq g(q \xi)j_2(q |\xv|) \\
I_4[g] & = \int_0^{\Lambda} dq g(q \xi)j_4(q |\xv|).
\end{align}
Thus,
\begin{align}
\mathcal{F} & = \mathcal{F}_A+\mathcal{F}_B-\mathcal{F}_C \\
& = \frac{\xi^\phi}{\pi} \Big\{ I_1[g_1] + P_2(\cos\theta_x)
I_2[g_2] \\
& \qquad \left[ \frac{1}{35}P_2(\cos\theta_x) + \frac{1}{8} \sin^2
\theta_x \cos 4 \phi_x \right] I_4 [g_C] ,
\end{align}
where $g_1 = g_A + \frac{2}{3}g_B - \frac{1}{15}g_C$ and $g_2 =
\frac{2}{3} g_B + \frac{2}{21}g_C$.  Thus we need only evaluate
the three integrals $I_1$, $I_2$, and $I_3$.

\subsubsection{$|\xv| \gg \xi >\Lambda^{-1}$ in Three Dimensions}

In this limit, in integrals with integrands proportional to $j_n
(q |\xv|)$, we set $y= q|\xv|$, replace $g(y \xi/|\xv|)$ by $g_0$
and replace the upper limit, $\Lambda |\xv|$, of integration by
$\infty$.  In the part of the integral $I_1$ not proportional to
$j_0(q |\xv|)$, we set $ y = q \xi$.  The result is
\begin{align}
I_1 &\sim g_0 \frac{\pi}{2}\left[\frac{2}{\pi} \int_0^{\Lambda
\xi}\frac{g(y)}{g_0} dy
- \frac{1}{|\xv|} \right]\\
I_2 & \sim \frac{g_0}{|\xv|} \int_0^{\infty} dy j_2 (y) =
\frac{g_0 \pi}{4 |\xv|} \\
I_3 & \sim \frac{g_0}{|\xv|} \int_0^{\infty} j_4 ( y) = \frac{3
g_0\pi}{16 |\xv|} .
\end{align}

\subsubsection{$\Lambda^{-1} \ll |\xv| \ll \xi$ in Three Dimensions}

To treat this limit, as in $2D$, we use the function $h(u)$ [Eq.\
( \ref{eq:h(u)})].  To evaluate $I_1$, we break up the limits of
integration in much the same way we did in $2D$.  The result is
\begin{align}
I_1 & = g_{\infty} \xi^{-\phi} |\xv|^{\phi -1} \\
&\,\, \times \Big\{ \int_0^1 dy
y^{-\phi} [1 - j_0(y)] - \int_1^{\infty} dy y^{-\phi} j_0(y) \\
& \qquad + \frac{1}{1-\phi}[(\Lambda|\xv|)^{1-\phi} -1] \\
& \qquad + \left(\frac{\xi}{|\xv|}\right)^{1-\phi}\int_0^{\Lambda
\xi} du u^{-\phi} [h(u) -1] \Big\}
\end{align}
for $0<\phi<3$.  The dominant behavior for $1<\phi<3$ and
$0<\phi<1$ is then
\begin{equation}
I_1 \sim
 \begin{cases}
 g_\infty \xi^{-\phi} |\xv|^{\phi-1} \mathcal{A}_3 (\phi) & \text{$1<\phi<3$} \\
 \frac{1}{\xi}\int_0^{\Lambda\xi} du g(u) -
 g_{\infty}\xi^{-\phi}|\xv|^{-(1-\phi)} \mathcal{C}_3(\phi) & \text{$0<\phi<1$}.
 \end{cases} ,
\end{equation}
where
\begin{subequations}
\begin{align}
\mathcal{A}_3(\phi) & = \int_0^{\infty}y^{-\phi}[1 - j_0 ( y)] \\
\mathcal{C}_3 ( \phi) & = \int_0^{\infty} y^{-\phi} j_0(y) .
\end{align}
\end{subequations}

\subsection{One Dimension}

In $1D$, there is only one function to evaluate
\begin{align}
\calF(x) & = 2 \xi^{\phi} \int_{- \Lambda}^{\Lambda} \frac{dq}{2
\pi} \frac{1}{q^2} g( q \xi) [ 1 - \cos (q |x|)] \nonumber\\
& = \frac{2 \xi^{\phi}}{\pi}\int_0^{\Lambda|x|} dy \frac{1}{y^2}
g( y \xi /|x|)(1 - \cos y) .
\end{align}
The limit $|x| \gg \xi$ is obtained as before by replacing $g( y
\xi/|x|)$ with $g_0$ and letting $\Lambda |x| \rightarrow \infty$:
\begin{equation}
\calF ( x ) = \frac{2}{\pi} |x| g_0 \xi^{\phi} \int_0^{\infty} dy
\frac{1- \cos y}{y^2} = g_0 \xi^{\phi} |x| .
\end{equation}

In the limit $\Lambda^{-1} \ll |x| \ll \xi$, we introduce $h(y)$
as in  $2D$ and $3D$: $\calF ( x) = (2 g_{\infty} |x|/\pi) \calK$,
where
\begin{align}
\calK & = &\int_0^{\Lambda |x|} dy h( y \xi /|x|) y^{-(2 + \phi)}
(1 - \cos y) \nonumber \\
& = &\calK_1 + \int_0^1 dy y^{-(2 + \phi)}[1 - \cos y -
(y^2/2)]\nonumber \\
& & + \int_1^{\infty} y^{-(2 + \phi)} [1 - \cos y] ,
\label{eq:calK}
\end{align}
where
\begin{eqnarray}
\calK_1 & = &\frac{1}{2} \int_0^1 dy y^{-\phi} h( y \xi/|x|)
\nonumber \\
& = &\frac{1}{2}\left(\frac{\xi}{|x|}\right)^{\phi -1} \Big\{
\frac{1}{1 - \phi} \Big(\left(\frac{\xi}{|x|}\right)^{1- \phi} -1
\Big] \label{eq:calK1} \\
& & + \int_0^2 du h(u) u^{-\phi} + \int_1^{\infty} du u^{-\phi}
[h(u) - 1]  \Big\}. \nonumber
\end{eqnarray}
Combining Eqs.\ (\ref{eq:calK}) with (\ref{eq:calK1}), we find
\begin{equation}
\calF(x) \sim
  \begin{cases}
  \frac{2}{\pi} g_{\infty} |x|^{1+ \phi} \mathcal{A}_1 ( \phi) &
  \text{if $\phi<1$} , \\
  \frac{1}{\pi} \ln \nu \xi/|x| & \text{if $\phi = 1$} ,\\
  \frac{1}{\pi} \xi^{\phi - 1} |x|^2 \mathcal{C}_2 (\phi) & \text{if
  $\phi >1$} ,
  \end{cases}
\end{equation}
where $\nu$ is given by Eq. ( \ref{eq:nu}) and $\mathcal{C}_2 (
\phi)$ is given by Eq.\ (\ref{eq:calC}) and where
\begin{equation}
\mathcal{A}_1 = \int_0^{\infty} \frac{1 - \cos y}{y^{2 + \phi}} .
\end{equation}

The $|\xv|\ll \xi$ limits of both $I_2$ and $I_3$ can be obtained
by simply by replacing $g(q \xi)$ by $(q\xi)^{-\phi}g_\infty$:
\begin{align}
I_2& \sim g_\infty \xi^{-\phi}|\xv|^{\phi-1} \int_0^{\infty}
dyy^{-\phi} j_2(y) & \text{$0<\phi<3$} \\
I_3 & \sim g_\infty \xi^{-\phi}|\xv|^{\phi-1} \int_0^{\infty}
dyy^{-\phi} j_4(y) & \text{$0<\phi<5$} .
\end{align}

\section{\label{App:D}Evaluation of $\cal{G}_{\omega} ( \xv )$}

In this appendix we will evaluate the rotational correlation
function $G_{\omega}( \xv)$ in two dimensions. To lowest order in
$\Delta^K$,
\begin{equation}
G_{\omega}( \qv ) = \frac{1}{4}\epsilon_{ri}\epsilon_{r'i'}q_r
q_{r'} \chi_{ip}^0(\qv) \chi_{i'p'}^0 ( - \qv)
\sum_{\alpha}S_{\alpha pp'} ,
\end{equation}
where $S_{\alpha pp'}$ is defined in Eq.\ (\ref{eq:Salphapp'}).
The product $\epsilon_{ri}\epsilon_{r'i'}$ is simply
$\delta_{rr'}\delta_{ii'}-\delta_{ri'}\delta_{ir'}$, and
$\epsilon_{ri}\epsilon_{r'i'}q_r q_r' = q^2 \delta_{ii'}^T$.  When
this operates on $\chi_{ip}^0\chi_{i'p'}^0$, it projects out the
transverse part leaving $\delta_{pp'}^T/(\mu^2 q^2)$. Thus
\begin{equation}
G_{\omega}( \qv ) = \frac{\gamma_{xy}^2}{\mu^2}
\sum_{\alpha}S_\alpha^T =
\frac{\gamma_{xy}^2}{\mu^2}(\Delta_A^{\omega} -
\Delta_{C}^{\omega} \hatq_x^2 \hatq_y^2) .
\end{equation}
where $\Delta_A^{\omega} = \Delta_3 + 4 \Delta_6 + 2 \Delta_7 + 2
\Delta_8$ and $\Delta_C^{\omega} = 4 \Delta_3 + 15 \Delta_6 + 4
\Delta_7$.

When $\Delta_{A,C}^{\omega} (\qv)$ have a Lorentizan form, we need
to evaluate two integrals to determine $G_{\omega}(\xv)$:
\begin{equation}
F_1(\xv) =\int \frac{d^2 q}{(2 \pi)^2} \frac{1}{q^2 +
\xi^{-2}}e^{i\qv\cdot \xv} = \frac{1}{2 \pi} K_0 ( |\xv|/\xi)
\end{equation}
and
\begin{eqnarray}
F_2( \xv ) & = & \int \frac{d^2 q}{(2 \pi)^2}\frac{\hatq_x^2
\hatq_y^2}{q^2 + \xi^{-2}}e^{i\qv\cdot \xv} \nonumber \\
& = & \frac{1}{2\pi} \int_0^{\infty}\frac{q d q}{q^2 +
\xi^{-2}}\int \frac{d \phi}{2 \pi} \cos^2 \phi \sin^2 \phi e^{i q
|\xv| \cos(\phi - \psi)} \nonumber \\
& = & \frac{1}{16 \pi} \Bigl[ K_0 ( |\xv|/\xi) \nonumber \\
& & - \cos 4 \psi \left(-\frac{48 \xi^4}{|\xv|^4} +
\frac{4\xi^2}{|\xv|^2} +K_4 ( |\xv|/\xi)\right)\Bigr] ,
\end{eqnarray}
where $\qv = q ( \cos \phi ,\sin \phi)$, $\xv = |\xv|( \cos \psi,
\sin \psi)$ and $K_n(y)$ is the Bessel function of imaginary
argument.

\bibliography{elasticity}

\begin{thebibliography}{36}
\bibitem{pressure} Though the focus of this paper is on elastic
networks composed of springs with a preferred length, our
formalism applies more generally.  It applies for example to
systems of particles interacting with repulsive forces only but
confined to a finite volume.  The container exerts and isotropic
pressure $p$, and there is an additional term in the Hamiltonian
equal to $-p (V - V_0) = - \int d^d x p \ln
(\Lambda_{ki}\Lambda_kj )^{1/2}$, where $V$ is volume in the
target space, $V_0$ is that in the reference space, and 
$\Lambda_{ij} = \partial R_i \partial x_j$.  Note that
this term is totally rotationally invariant and depends only on
$u_{ij}$ Minimization with respect to $\Lambda_{ij}$ produces an
equilibrium $\Lambda_{ij}^0$ that depends on $p$ and produces a
new reference state with spatial positions $x'_i =
\Lambda_{ij}^0 x_j$.
\bibitem{rot_inv} Though it is clear by construction that the
stress tensor term in Eq.\ (\ref{eq:stress3}) is rotationally
invariant, it is instructive to derive this explicitly.  Under a
rigid rotation described by a rotation matrix $U_{ij}$, the
displacement field transforms according to $u_i \rightarrow
(U_{ij} - \delta_{ij}) x_j + U_{ij} u_j$, and $\partial_j u_k
\partial_j u_k \rightarrow 2 \delta_{ij} - U_{ij} - U_{ji} + 2
\partial_j[(U_{ip} - \delta_{ip})u_p] + \partial_i u_k \partial_j
u_k$.  The contribution of the constant term to $\int d^d x
\tilde{\sigma}_{ij} \partial_i u_k \partial_j u_k$ vanishes because
the spatial average of $\tilde{\sigma}_{ij}$ is zero, and the
contribution from the terms with a single gradient vanish because
$\int d^d x \tilde{\sigma}_{ij} \partial_
j h_i ( \xv)$ vanishes
for any function $h_i$.
\end{thebibliography}

\end{document}